\documentclass{elsarticle}
\usepackage[utf8]{inputenc}
\usepackage{graphicx}
\usepackage{amsfonts}
\usepackage{amssymb}
\usepackage{mathtools}
\usepackage{subcaption}
\usepackage{xcolor}
\journal{ }
\newcommand{\bsym}[1]{\boldsymbol{#1}}

\title{Li$_x$CoO$_2$ phase stability studied by machine learning-enabled scale bridging between electronic structure, statistical mechanics and phase field theories}

\author[umme]{G.H. Teichert}
\author[umme]{S. Das}
\author[tri]{M. Aykol}
\author[tri]{C. Gopal}
\author[umme,umms,micde]{V. Gavini}
\author[umme,umm,micde]{K. Garikipati\corref{mycorrespondingauthor}}
\ead{krishna@umich.edu}
\cortext[mycorrespondingauthor]{Corresponding Author}

\address[umme]{Department of Mechanical Engineering, University of Michigan}
\address[tri]{Toyota Research Institute, Los Altos, CA}
\address[umms]{Department of Materials Science \& Engineering, University of Michigan}
\address[umm]{Department of Mathematics, University of Michigan}
\address[micde]{Michigan Institute for Computational Discovery \& Engineering, University of Michigan}

\begin{document}

\begin{abstract}
Li$_x$\textit{TM}O$_2$ (TM={Ni, Co, Mn}) are promising cathodes for Li-ion batteries, whose electrochemical cycling performance is strongly governed by crystal structure and phase stability as a function of Li content at the atomistic scale. Here, we use Li$_x$CoO$_2$ (LCO) as a model system to benchmark a scale-bridging framework that combines density functional theory (DFT) calculations at the atomistic scale with phase field modeling at the continuum scale to understand the impact of phase stability on microstructure evolution. This scale bridging is accomplished by incorporating traditional statistical mechanics methods with integrable deep neural networks, which allows formation energies for specific atomic configurations to be coarse-grained and incorporated in a neural network description of the free energy of the material. The resulting realistic free energy functions enable atomistically informed phase-field simulations. These computational results allow us to make connections to experimental work on LCO cathode  degradation as a function of temperature, morphology and particle size.
% \cg{Insert 1-2 sentences on the phase field results}
\end{abstract}

\maketitle

\section{Introduction}
Layered oxides of type Li$_x$\textit{TM}O$_2$ (TM={Ni, Co, Mn}) are commonly used cathodes in Li-ion batteries. Key to their versatility is the ability to tune the chemistry specific to an application. While LiCoO$_2$ (LCO) has traditionally been the cathode of choice for consumer electronics, Ni-rich compositions are attractive candidates for high energy density batteries, which are increasingly in demand with electrification of transportation. However, the cycling performance of Ni-rich NMC cathodes is known to suffer from surface phase transitions \cite{Zheng2019,Mu2018}, mechanical failures \cite{Yang2019,Mao2019}, and undesired degradation reactions with electrolytes \cite{Liu2015,Hou2017}.
A common strategy to improve the cycling and structural stability of layered oxides is extrinsic doping with trace amounts of foreign elements \cite{Kim2017,Kong2019,Weigel2019}.
Understanding the defect-thermodynamics of dopants and how that impacts the crystal structure stability and  long-term electrochemical properties of NMCs from first principles requires a modeling framework that span lengthscales ranging from the atomistic to the continuum. However, scale-bridging is inherently a hard problem to address and prior studies, as we outline below, have looked at the different length-scales in isolation. Here, we benchmark a recently published scale-bridging framework \cite{Teichert2019,Teichert2020} against Li$_x$CoO$_2$ (LCO), a well-characterized model cathode, by computing its thermodynamic phase behavior and microstructure evolution under Li intercalation.

The rich phase-behavior of LCO has been extensively studied by experiments. A first order phase transition has been identified for compositions $0.75 \leq x \leq 0.94$, caused by a metal-insulator transition \cite{Reimers1992,Menetrier1999}. Reimers and Dahn \cite{Reimers1992} reported on an ordering that appears at $x = 0.5$, confirmed to be a row ordering \cite{ShaoHorn2003,Takahashi2007}.
Shao-Horn et al. \cite{ShaoHorn2003} found evidence of ordering at $x = 0.33$ at low temperatures ($\sim$100 K).
Additionally, charge ordering Co$^{3+}$ and Co$^{4+}$ atoms at compositions of $x = 0.5$ and $x = 0.67$ was observed at a temperature of 175 K, while Co$^{3+}$ and Co$^{4+}$ occupation was random at room temperature \cite{Motohashi2009,Takahashi2007}. Multiple  studies have also identified a transition from the O3 host structure to the H1-3 structure at compositions below $x = 0.33$ \cite{Chen2004,Chang2013}.

Alongside the experimental work have been the numerous computational studies that have also explored LCO. \emph{Ab initio} thermodynamic approaches combining density functional theory (DFT) calculations and statistical mechanics \cite{VanderVen1998, Wolverton1998} have been used to systematically predict phase diagrams, free energy data, and voltage curves. The first-principles methods also have been combined with experimental results and CALPHAD models to define functions for the Gibbs energy \cite{Abe2011,Chang2013}.
However, quantities computed from the atomistic and lattice models have not informed the more coarse-grained methods, such as phase field models to study the thermodynamics and kinetics of phase transitions in these systems. Recently, a continuum level phase field approach has combined free energy and misfit strain information from experiments with the predicted diffusivity from first-principles calculations to model the metal-insulator phase transition that has been seen experimentally \cite{Nadkarni2019}. In effect, while the  link between \emph{ab initio} and statistical mechanics computations is well established, there remains an absence, so far, of comprehensive scale bridging frameworks.

\begin{figure}[tb]
        \centering
	\includegraphics[width=0.8\textwidth]{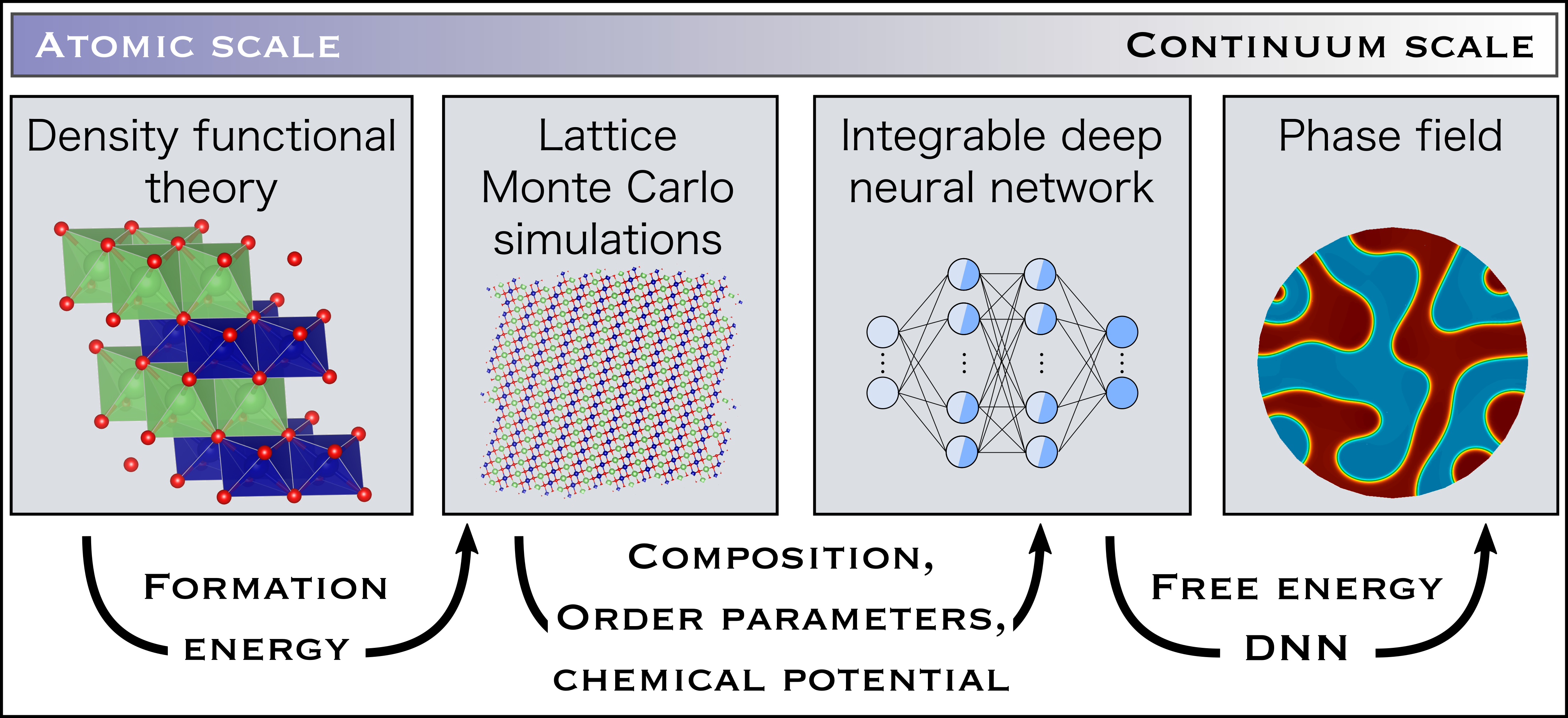}
\caption{Flowchart outlining the data and methods used to bridge from the atomic scale up to the continuum scale.}
\label{fig:flowchart}
\end{figure}

The current work seeks to demonstrate a systematic framework combining some of the computational approaches described above with advances in machine learning to bridge electronic structure calculations, statistical mechanics approaches and continuum scale simulations, Figure \ref{fig:flowchart}. Our use of O3 layered LCO to benchmark the approach is motivated by the observation that, to the best of our knowledge, this path to scale bridging has not been presented for this cathode material. 

We perform DFT$+U$ calculations using the well-known Hubbard correction~\cite{Cococcioni2005} to capture  electronic correlation in transition metal oxides with localized $d$ orbitals. Combined with van der Waals functionals, this ensures closer agreement with experimental voltages in layered LCO~\cite{Aykol2015}. A cluster expansion Hamiltonian is parameterized by the DFT and used in Monte Carlo sampling to compute the point defect thermodynamics as a function of temperature and chemical potential, and to construct temperature - composition phase diagrams \cite{PhysRevB.86.134117, y2002first}.

For bridging statistical mechanics and continuum representations of the free energy density function (and its derivatives), we turn to integrable deep neural networks (IDNNs) introduced previously \cite{Teichert2019,Teichert2020}. IDNNs can be trained to derivative data, e.g. chemical potential data generated through Monte Carlo sampling, and analytically integrated to recover the antiderivative function, e.g. the free energy function needed in phase field models. The chemical potentials and free energies thus obtained are used to confirm a match with extensively reported phase diagrams for LCO, as well as voltage curves over the composition range of $[0.33,1]$. Finally, using the obtained free energies, we carry out phase field studies that take a step beyond the construction of equilibrium phase diagrams by accessing dynamics of phase formations and particle morphologies. Specifically, we study spinodal decomposition as well as charge-discharge cycles on single LCO particles at temperatures of practical interest, and for different particle sizes. These computational results allow us to make connections to experimental work on LCO cathode  degradation as a function of temperature, morphology and particle size. By this, they allow us to address experimental and technologically relevant scenarios. They also serve as a culminating demonstration of predictive computations enabled by our scale bridging framework.  

The paper is laid out as follows: Section \ref{sec:methods} sets forth the methods used in the paper consisting of the DFT calculations, statistical mechanics sampling, IDNN setup and training, and the phase field equations. Section \ref{sec:results} describes and discusses the results of each step of the framework while making comparisons with experimental observations. Discussion and conclusions are presented in Section \ref{sec:conclusions}. 

\section{Methods}
\label{sec:methods}
In this section we detail the methods that were previewed in the introduction, starting with the DFT calculations and moving on to the statistical mechanics simulations that they inform. This is followed by the machine learning techniques, specifically IDNNs \cite{Teichert2019,Teichert2020} that enable bridging of the atomic to continuum scales via free energy density representations. This provides a principled basis for continuum simulations of the dynamics of LCO particle morphology via phase field models whose computational treatment is the culmination of the scale bridging framework (Figure \ref{fig:flowchart}).

\subsection{Density functional theory methods}
\label{sec:DFT}

The formation energy $E$ of a configuration with Li composition $x$ can be calculated using total energy values computed using density functional theory (DFT) according to the following equation \cite{VanderVen1998}:
\begin{align}
    E &= E^\mathrm{tot} - xE^\mathrm{tot}_{\mathrm{LiCoO}_2} - (1 - x)E^\mathrm{tot}_{\mathrm{CoO}_2}
    \label{eqn:form_energy}
\end{align}
where $E^\mathrm{tot}$, $E^\mathrm{tot}_{\mathrm{LiCoO}_2}$, and $E^\mathrm{tot}_{\mathrm{CoO}_2}$ are the total energies for the given configuration, LiCoO$_2$, and CoO$_2$, respectively. In performing DFT calculations for Li$_x$CoO$_2$, Aykol and co-authors found that it is important to include not only the Hubbard correction (DFT+$U$), which reduces self-interaction errors \cite{Anisimov1991,Zhou2004,Aykol2014}), but to also include the van der Waals (vdW) interactions \cite{Aykol2015}. The effect of the vdW interactions allows the voltage predicted by DFT to match the experimental voltage when using an appropriately tuned value of $U$. While both vdW-corrections and vdW-density functionals (vdW-DF) improve the DFT results, predictions are most consistent and accurate with vdW-DF. 

\begin{figure}[tb]
        \centering
    \begin{minipage}[t]{0.25\textwidth}
    \centering
	\includegraphics[width=0.5\textwidth]{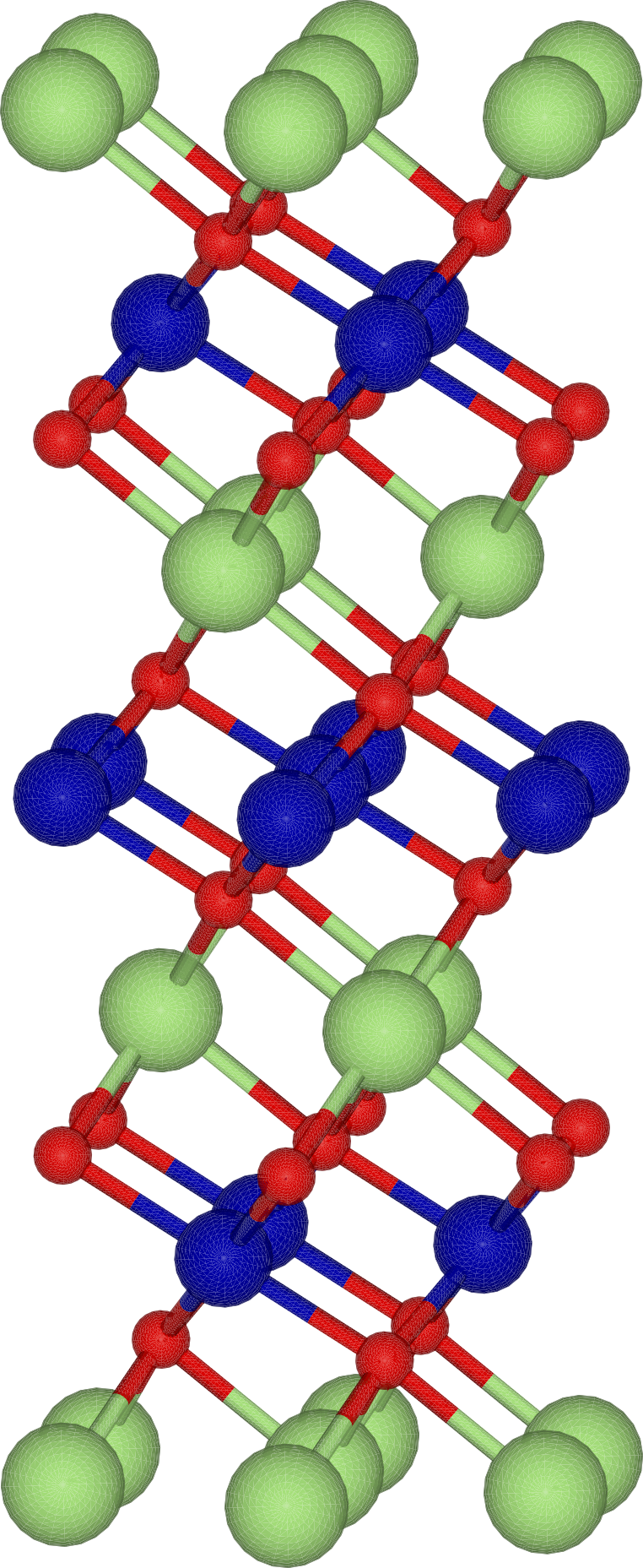}
	\subcaption{Ball-and-stick}
    \end{minipage}%
    \begin{minipage}[t]{0.4\textwidth}
        \centering
	\includegraphics[width=0.9\textwidth]{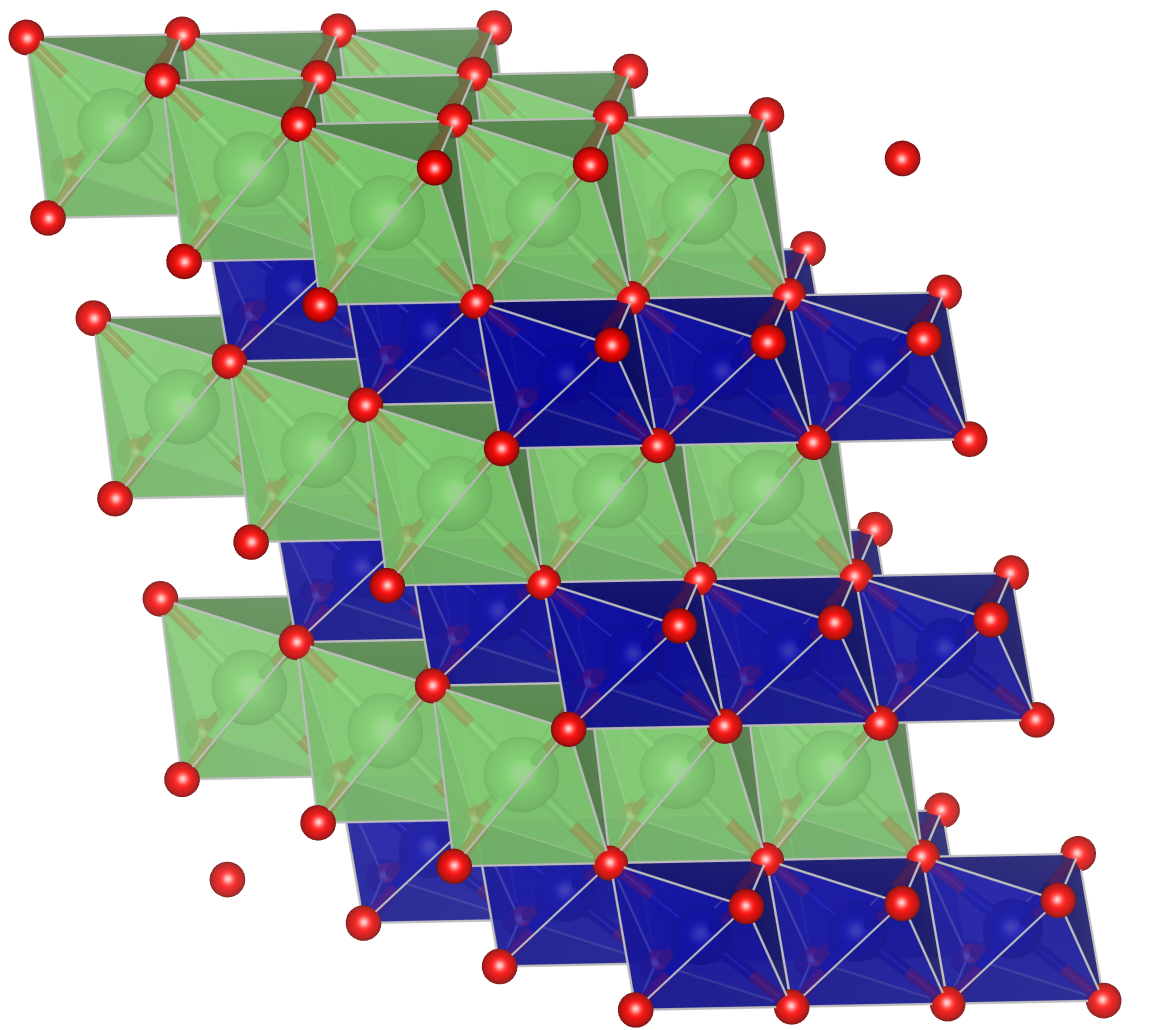}
	\subcaption{Polyhedral}
\end{minipage}
\caption{Two representations of the crystal structure of layered O3 LiCoO$_2$, where the Li atoms are green, Co are blue, and O are red (visuals generated using VESTA \cite{Momma2011}).}
\label{fig:crystal_struc}
\end{figure}

In this workflow, we use a simplified rotational-invariant formulation of DFT+$U$~\cite{Cococcioni2005} and a vdW-DF exchange correlation functional, namely the optB88 exchange correlation functional \cite{Thonhauser2007,Klime2009,Langreth2009,Sabatini2012,Aykol2015,Thonhauser2015,Berland2015}, to calculate the formation energy on a chosen subset of LCO configurations. We perform the ground-state DFT calculations with geometry optimization in \texttt{Quantum Espresso} \cite{QE-2009,QE-2017}, using projector augmented-wave (PAW) pseudopotentials calculated with the Perdew-Burke-Ernzerhof (PBE) functional from \texttt{PSlibrary 1.0.0} \cite{dalcorso2014,pslibrary}. The values for the wave function and charge density cutoffs are chosen in two steps. Starting with the cutoff values suggested in the pseudopotential file for Co, we first increase only the charge density cutoff until the total energy converges to $<$1 meV/atom. Next, we increase both the wave function and the charge density cutoffs, maintaining the ratio of the two values, until the total energy again converges to $<$1 meV/atom, giving a wave function cutoff of 55 Ha and a charge density cutoff of 301.5 Ha. A k-point grid of $6 \times 6 \times 3$ is also used to ensure total energy convergence within $<$1 meV/atom. Structural optimization is performed until cell stress and ionic forces are under 0.5 kbar and 0.00005 Ha/Bohr, respectively. The crystal structure for LiCoO$_2$ is shown in Figure \ref{fig:crystal_struc}. 

\begin{figure}[t]
    \centering
    \includegraphics[width=0.8\textwidth]{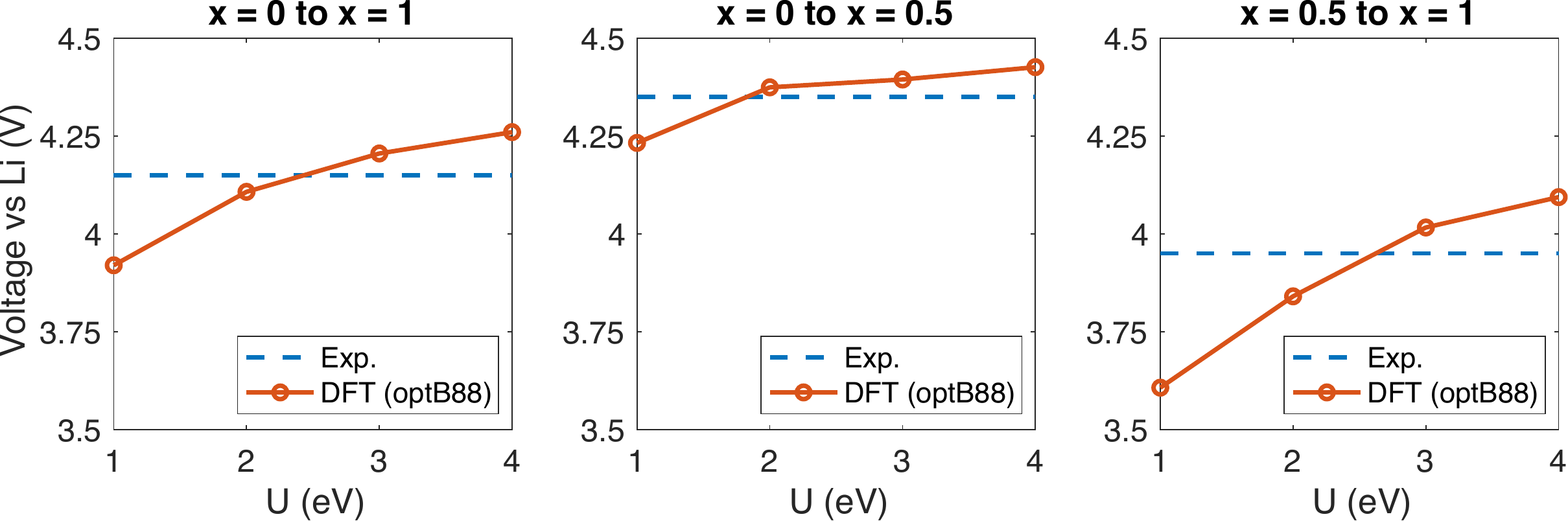}
    \caption{Calculated voltage across various compositions as a function of U, compared with experimental voltage \cite{Aykol2015}.}
    \label{fig:volt_U}
\end{figure}

In this work, we calibrate the Hubbard $U$ parameter to match experimental average lithiation voltages over various ranges of Li composition. To do this, we compute the voltage at increasing $U$ for given composition values of $x_1$ and $x_2$ \cite{Meredig2010}, using crystal structures previously reported from experiments. The average voltage from $x_1$ to $x_2$ is calculated using the following equation \cite{Aydinol1997,Aykol2015}:
\begin{align}
    V = -\frac{E[\mathrm{Li}_{x_2}\mathrm{CoO}_2] - E[\mathrm{Li}_{x_1}\mathrm{CoO}_2] - (x_2 - x_1)E[\mathrm{Li}]}{(x_2 - x_1)e}
\end{align}
where $E$ is the calculated total energy from DFT and $e$ is the elementary charge. We compute the average voltage over the following ranges of Li composition: $x\in[0,1]$, $x\in[0,1/2]$, and $x\in[1/2,1]$. By comparing with experimental voltages (Figure \ref{fig:volt_U}), we select an appropriate value of $U$.

\subsection{Statistical mechanics}
\label{sec:statmech}

In this work, we follow the statistical mechanics approach outlined by Van der Ven, et al. \cite{VanderVen1998} in their first-principles study of Li$_x$CoO$_2$ (LCO), and of Natarajan, et al. \cite{Natarajan2017}, among others.

\subsubsection{Cluster expansion for formation energy}
Given the expense of DFT calculations, we adopt cluster expansions to access the large numbers of configurations needed in the statistical mechanics studies. The formation energy is written as $E(\bsym{\sigma})$, where  $\bsym{\sigma}$ is the configuration vector with the occupancy variable $\sigma_i = 1$ if Li occupies the site $i$ and $\sigma_i = 0$ if the site contains a vacancy. As is common, we define $E(\bsym{\sigma})$ by computing the formation energy with DFT for a subset of configurations and use these values to parameterize a cluster expansion \cite{Sanchez1984,deFontaine1994} as a rapidly queryable surrogate for the formation energy. We use the \texttt{CASM} software\cite{casm}, which facilitates the construction and parameterization of cluster expansion Hamiltonians and their use in Monte Carlo simulations, to select configurations for the DFT computations and perform the statistical mechanics calculations in this work \cite{VanderVen2010,thomas2013,puchala2013}. Candidate configurations are chosen for O3-Li$_x$CoO$_2$ with Li compositions from $x=0$ to $x=1$. 

A cluster is a collection of potential Li sites. Given cluster $\alpha$ of sites $\alpha = \{i,j,\dots,k\}$, a polynomial $\phi_\alpha$ can be defined as the product of occupancy variables of those sites, i.e. $\phi_\alpha = \sigma_i\sigma_j\cdots\sigma_k$. A cluster expansion is a linear combination of the polynomials $\phi_\alpha$, leading to the following form for the formation energy:
\begin{align}
    E(\bsym{\sigma}) &= V_0 + \sum_\alpha V_\alpha \phi_\alpha(\bsym{\sigma}),
\end{align}
where the coefficients $V_0$ and $V_\alpha$ are optimized for a ``best fit'', and are called effective cluster interactions (ECI).

We fit a cluster expansion to the formation energy calculated using DFT for 333 configurations for the O3 host structure using the \texttt{CASM} code. A sparse regression technique that combines a genetic algorithm with weighted linear regression is used to perform the fit. In this approach, a number of candidate basis functions are created using pairs, triplets, etc. of lattice sites that are within a given distance. The genetic algorithm is used to select a subset of these basis functions to include in the cluster expansion. The coefficients of each subset of basis functions are calculated using linear regression. We use cluster types up to quadruplets, with maximum lattice site distances shown in Table \ref{tab:clex_basis}, for a total of 221 candidate basis functions. During the linear regression, we apply weights to bias the fit towards greater accuracy for configurations on or near the convex hull. Additionally, the cluster expansion is selected such that the convex hull constructed from the cluster expansion predictions for all 333 configurations consists of the same configurations as in the DFT convex hull.

\begin{table}[t]
    \caption{Basis function specifications used for cluster expansion, resulting in a total of 221 candidate basis functions.}
    \centering
    \begin{tabular}{| c | c |}
    \hline
    cluster type & max. distance (\AA)\\
    \hline
    pairs & 24\\
    triplets & 8\\
    quadruplets & 6\\
    \hline
    \end{tabular}
    \label{tab:clex_basis}
\end{table}

\subsubsection{Monte Carlo sampling}
Given $E(\bsym{\sigma})$, we sample within the semi-grand canonical ensemble, in which the chemical potential is specified and the corresponding composition and/or order parameters are determined through ensemble averaging. The partition function for the semi-grand canonical ensemble is the following:
\begin{align}
    \Theta &= \sum_{\bsym{\sigma}} \exp{\left(-\frac{E(\bsym{\sigma}) - M\left[x(\bsym{\sigma})\mu_0 + \bsym{\eta(\bsym{\sigma})}\cdot\bsym{\mu}\right]}{k_B T}\right)} \label{eqn:part2}
\end{align}
where $\bsym{\eta}$ represents the vector of order parameters explained below, $\mu_0$ is the chemical potential associated with composition, $\bsym{\mu}$ is the vector of chemical potentials associated with the order parameters, $M$ is the number of reference supercells that tile the configuration, $k_B$ is the Boltzmann constant, and $T$ is the temperature. Each chemical potential is the derivative of the free energy with respect to its corresponding composition or order parameter.

The chemical potential associated with Li composition is related to the voltage as a function of composition, $V(x)$ as
\begin{equation}
    V(x) = -\frac{\mu_\mathrm{Li}^\mathrm{cathode}(x) - \mu_\mathrm{Li}^\mathrm{anode}}{e},
    \label{eqn:voltage}
\end{equation}
where $\mu_\mathrm{Li}^\mathrm{cathode}$ is the chemical potential with respect to Li for the cathode (LCO) and $\mu_\mathrm{Li}^\mathrm{anode}$ is the chemical potential of the anode. Taking the anode to be Li metal, $\mu_\mathrm{Li}^\mathrm{anode}$ is a constant equal to the Gibbs free energy of Li \cite{Aydinol1997}. It is necessary for the two chemical potentials to have a consistent reference, however. We approximate the Gibbs free energy of Li with the total energy calculated using DFT, but the chemical potential for LCO reported from Monte Carlo sampling is based on formation energies that were re-referenced to the end members, according to the linear relation in Eq. (\ref{eqn:form_energy}). Since the chemical potential is the derivative of the free energy with respect to composition, this re-referencing shifted the chemical potential of LCO by a constant $k$ equal to the derivative of Eq. (\ref{eqn:form_energy}) with respect to $x$, i.e. $k := -E^\mathrm{tot}_{\mathrm{LiCoO}_2} + E^\mathrm{tot}_{\mathrm{CoO}_2}$. Therefore, we reverse the shift in the chemical potential by subtracting $k$ from $\mu_\mathrm{Li}^\mathrm{cathode}(x)$ before computing the voltage according to Eq. (\ref{eqn:voltage}). This imposes a consistent reference for both $\mu_\mathrm{Li}^\mathrm{cathode}$ and $\mu_\mathrm{Li}^\mathrm{anode}$.

A limitation with Monte Carlo sampling within the semi-grand canonical ensemble is that it does not produce data within the unstable regions associated with phase separation. While this is sufficient for producing phase diagrams and predicting voltage, phase field simulations require free energy information within these two-phase regions in order to consistently resolve phase interfaces. For this reason, we also employ umbrella sampling methods, in which bias potentials are applied to sample within the unstable regions that would otherwise be missed \cite{Natarajan2017,mishin2004,Sadigh2012a,Sadigh2012b}. The partition function used in this case, then, is the following:
\begin{align}
    \Theta &= \sum_{\bsym{\sigma}} \exp{\left(-\frac{E(\bsym{\sigma}) + M\left[\phi_0(x(\bsym{\sigma}) - \kappa_0)^2 + \sum_{i=1}^n\phi_i(\eta_i(\bsym{\sigma}) - \kappa_i)^2\right]}{k_B T}\right)} \label{eqn:part3}
\end{align}
where $\phi_i$ and $\kappa_i$ determine the curvature and center of the bias potential, respectively, and $n$ is the number of order parameters. Usually, $\phi_i$ can be fixed at an appropriate value while $\kappa_i$ is varied to sample across the desired composition and order parameter space within the Monte Carlo sampling routine. The ensemble average of the composition $\langle x\rangle$ and each order parameter $\langle\eta_i\rangle$ is related to its corresponding chemical potential through the bias parameters:
\begin{align}
    \frac{1}{M}\mu_0\Big|_{\langle x\rangle,\langle\bsym{\eta}\rangle} &= -2\phi_0(\langle x\rangle - \kappa_0)\\
    \frac{1}{M}\mu_i\Big|_{\langle x\rangle,\langle\bsym{\eta}\rangle} &= -2\phi_i(\langle\eta_i\rangle - \kappa_i), \qquad i=1,\ldots,n
\end{align}

\subsubsection{Symmetry-adapted order parameters}
\label{sec:order params}
In addition to composition, symmetry-adapted order parameters are useful for observing and tracking order-disorder transitions, as well as for identifying the various translational and rotational variants of a given ordering. The method for identifying symmetry-adapted order parameters is laid out elsewhere; e.g., see Natarajan, et al. \cite{Natarajan2017}. In this process, a supercell capable of representing all variants of the ordering is found, called a mutually commensurate supercell. The sublattice compositions of this supercell form a basis that can describe each variant as a vector $\bsym{x}$, where each component is 1 if the corresponding sublattice sites are fully occupied by Li and 0 otherwise. A matrix $\bsym{Q}$ that transforms the sublattice compositions $\bsym{x}$ to a vector of order parameters $\bsym{\eta}$ is constructed following an algorithm described by Thomas and Van der Ven \cite{thomas2017}.

The algorithm involves enumerating the group $P$ of matrices that transform between all variants of the given ordering. A matrix is formed, defined by its invariance to all transformation matrices in $P$. The eigenvalue decomposition of this $P$-invariant matrix is used to construct $\bsym{Q}$. While Thomas and Van der Ven deal with a general system, in simpler cases each nonzero eigenvalue is associated with either the average composition or the set of variants for a particular ordering. The eigenvectors form the rows of the matrix $\bsym{Q}$, defining the transformation from sublattice compositions to order parameters. In some cases, $\bsym{Q}$ maps to more order parameters than are necessary to describe a single ordering. When that occurs, a subset of rows can be extracted from $\bsym{Q}$ to create a reduced set of relevant order parameters. This algorithmic process is easily understood via an example, which is described in section \ref{sec:order parameters}.

\subsection{Integrable deep neural networks for free energy representations}
\label{sec:IDNN}

The IDNN representation is obtained for the free energy density function by training on derivative data in the form of pairs of chemical potential--the label--and corresponding composition or order parameter--the features \cite{Teichert2019,Teichert2020}. The integrability of the IDNN is built into its structure by constructing it as the derivative form of a standard fully-connected deep neural network (DNN). While it is straightforward to differentiate the equations describing the standard DNN to derive the equations for the IDNN, modern deep learning libraries make this step unnecessary. The user can simply construct a standard DNN and apply a gradient operation to create the IDNN, which is then used for training. Mathematically, a DNN can be denoted by a function $\bsym{Y}(\bsym{X},\bsym{W},\bsym{b})$ representing the chemical potentials $\langle \mu_i\rangle$, with arguments or inputs $\bsym{X}$ representing the composition $\langle x \rangle$, order parameters $\langle\eta_i\rangle$, weights $\bsym{W}$, and biases $\bsym{b}$. Training the DNN involves an optimization problem for the weights and biases, given the dataset $\{(\bsym{\widehat{X}}_\theta,\widehat{Y}_{\theta_k})\}$:
\begin{align}
    \bsym{\widehat{W}},\bsym{\widehat{b}} = \underset{\bsym{W},\bsym{b}}{\mathrm{arg\,min}}\,\mathrm{MSE}\left(\bsym{Y}(\bsym{X},\bsym{W},\bsym{b})\Big |_{\bsym{\widehat{X}}_\theta},\widehat{Y}_{\theta}\right)
\end{align}
The case of interest here is when the dataset $\{(\bsym{\widehat{X}}_\theta,\widehat{Y}_{\theta_k})\}$ is not available. Instead, we have the derivative dataset $\{(\bsym{\widehat{X}}_\theta,\widehat{y}_{\theta_k})\}$, where $\widehat{y}_{\theta_k}$ corresponds to the $k$-th partial derivative of $\widehat{Y}_{\theta}$. To use these data, the IDNN is defined as the gradient of $Y$ with respect to its inputs $\bsym{X}$, i.e. $\partial\bsym{Y}(\bsym{X},\bsym{W},\bsym{b})/\partial X_k$. The training is defined as follows:
\begin{align}
    \bsym{\widehat{W}},\bsym{\widehat{b}} = \underset{\bsym{W},\bsym{b}}{\mathrm{arg\,min}}\,\sum_{k=1}^n\mathrm{MSE}\left(\frac{\partial\bsym{Y}(\bsym{X},\bsym{W},\bsym{b})}{\partial X_k}\Big |_{\bsym{\widehat{X}}_\theta},\widehat{y}_{\theta_k}\right)
\end{align}
The resulting optimized weights $\bsym{\widehat{W}}$ and biases $\bsym{\widehat{b}}$ can be used with the function $\partial\bsym{Y}(\bsym{X},\bsym{\widehat{W}},\bsym{\widehat{b}})/\partial X_k$ to return a prediction of the chemical potential. Its antiderivative is exactly represented by using the same weights and biases in the function $\bsym{Y}(\bsym{X},\bsym{\widehat{W}},\bsym{\widehat{b}})$. For the current work, $\partial\bsym{Y}(\bsym{X},\bsym{\widehat{W}},\bsym{\widehat{b}})/\partial X_k$ gives the IDNN representation of the chemical potentials, and $\bsym{Y}(\bsym{X},\bsym{\widehat{W}},\bsym{\widehat{b}})$ is the DNN representation of the free energy.

We use the \texttt{Keras} and \texttt{Tensorflow} libraries to train IDNNs using the chemical potential and composition data for temperatures of 260 K, 300 K, and 340 K. The IDNN used consists of two hidden layers. For our initial studies we have used one input (Li composition) and one output (chemical potential with respect to composition). The IDNNs for 260 K, 300 K, and 340 K are constructed with 100, 60, and 20 neurons per layer, respectively, arrived at by hyperparameter optimization, due to differing complexity in the data at different temperatures. The hyberbolic tangent is used as the activation function, since we found it to better capture the small fluctuations in chemical potential around $x=0.5$ than the softplus function used by us in previous work \cite{Teichert2019,Teichert2020}. The IDNNs are trained for 50,000 epochs using the RMSprop optimizer and a batch size of 20. An initial learning rate of 0.002 is used, with periodic reductions in the learning rate applied during training. The training sets consist of 90, 85, and 69 data points for 260 K, 300 K, and 340 K, respectively.

\subsection{Phase field theory and associated computational framework}
\label{sec:phasefield}

The evolution of microstructure and phase changes can be modeled using the phase field equations. The Cahn-Hilliard equation \cite{CahnHilliard1958} models the dynamics of conserved quantities, such as composition, while nonconserved order parameter fields are modeled using the Allen-Cahn equation \cite{Allen1979}. When neglecting elastic effects, the total free energy of the system with $n$ order parameters can be described as follows:
\begin{align}
    \Pi[x,\bsym{\eta}] = \int\limits_\Omega \left(f(x,\bsym{\eta}) + \frac{1}{2}\chi_0|\nabla x|^2 + \sum_{i=1}^n\frac{1}{2}\chi_i|\nabla\eta_i|^2\right)\,\mathrm{d}V
\end{align}
where $\chi_i$ are the gradient parameters, and $f(x,\bsym{\eta})$ is the free energy density, to be represented by an analytically integrated DNN in this work.

The chemical potentials $\widetilde{\mu}_i$ used in the phase field equations are given by the variational derivatives of the total free energy, such that $\widetilde{\mu}_0 := \delta\Pi/\delta x$ and $\widetilde{\mu}_i := \delta\Pi/\delta\eta_i$, $i = 1,\ldots,n$:
\begin{align}
    \widetilde{\mu}_0 &= \frac{\partial f}{\partial x} - \chi_0 \nabla^2 x\\
    \widetilde{\mu}_i &= \frac{\partial f}{\partial \eta_i} - \chi_i \nabla^2 \eta_i, \qquad i=1,\ldots,n
\end{align}
The Cahn-Hilliard and Allen-Cahn equations, respectively, are the following:
\begin{align}
    \frac{\partial x}{\partial t} &= \nabla\cdot(\widetilde{M}\nabla\widetilde{\mu}_0) \label{eqn:CH}\\
    \frac{\partial \eta_i}{\partial t} &= -L\widetilde{\mu}_i, \qquad i=1,\ldots,n
    \label{eqn:AC}
\end{align}
where $\widetilde{M}$ is the mobility and $L$ is a kinetic coefficient. We substitute in the equations for the chemical potentials and write the governing equations in weak form to be solved using a mixed finite element method:
\begin{align}
    0 &= \int_\Omega \left(w_x\frac{\partial x}{\partial t} + \widetilde{M}\nabla w_x\cdot\nabla\tilde{\mu}_0\right)\mathrm{d}V - \int_{\partial\Omega}w_xj_n\mathrm{d}S\\
    0 &= \int_\Omega \left[w_{\tilde{\mu}_0}\left(\tilde{\mu}_0 - \frac{\partial f}{\partial x}\right) - \chi_0\nabla w_{\tilde{\mu}_0}\cdot\nabla x\right]\mathrm{d}V\\
    % 0 &= \int_\Omega \left[w_0\frac{\partial x}{\partial t} + M\left(\nabla w_0\cdot\nabla g_{,x}+\chi_0\nabla^2w_0\nabla^2 x\right)\right]\mathrm{d}V\\
    0 &= \int_\Omega \left[w_i\frac{\partial \eta_i}{\partial t} + L\left(w_i\frac{\partial f}{\partial \eta_i} + \chi_i\nabla w_i\cdot\nabla\eta_i\right)\right]\mathrm{d}V, \qquad i=1,\ldots,n
\end{align}
where $w_x$, $w_{\tilde{\mu}_0}$, and $w_i$ are weighting functions. For the equations written in this mixed formulation, the following Neumann boundary conditions have been applied to $x$, $\tilde{\mu}_0$, and $\eta_i$, $i = 1,\ldots,n$,  on $\partial\Omega$, where $\bsym{n}$ is the outward unit normal and $j_n$ is an influx:
\begin{align}
    \nabla x\cdot\bsym{n} &= 0\label{eq:dirbc-c}\\
    \widetilde{M}\nabla \tilde{\mu}_0\cdot\bsym{n} &= j_n\label{eq:neumbc-mu0}\\
    \nabla \eta_i\cdot\bsym{n} &= 0, \qquad i=1,\ldots,n\label{eq:dirbc-eta}
\end{align}

\begin{figure}[t]
    \centering
    \includegraphics[width=0.4\textwidth]{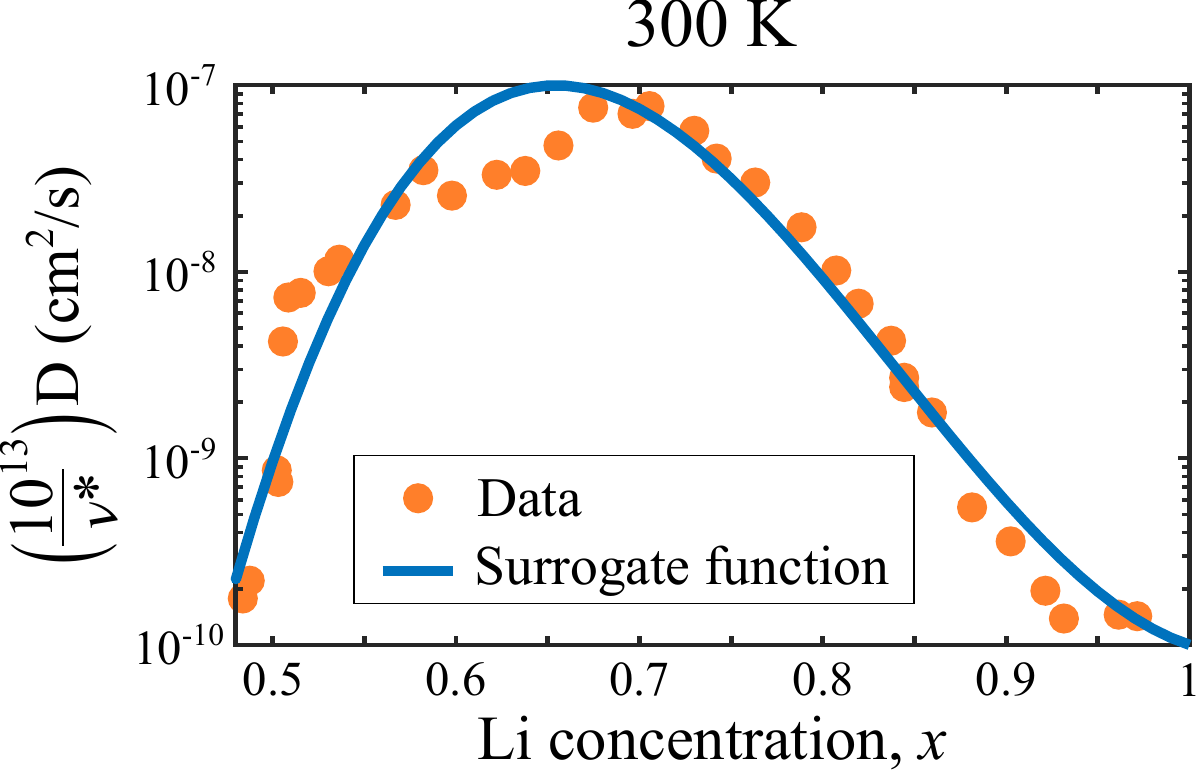}
    \caption{Surrogate function for the diffusivity of LCO in the Li composition range $x \in [0.5,1]$, plotted against first-principles diffusivity data of LCO at 300 K from Ref. \cite{VanderVen2000}.}
    \label{fig:diffusivity}
\end{figure}

To compare the behavior of LCO at different temperatures, we perform phase field simulations at 260 K, 300 K, and 340 K, informed by data from the Monte Carlo computations at these temperatures. The analytically integrated free energy DNNs for each of the three temperatures are used. We define the following composition-dependent surrogate function for the diffusivity $D$ at 300 K, using the predicted values from Ref. \cite{VanderVen2000}:
\begin{equation}
D = 0.01\exp(-274(1.05-x)(0.47-x)(1.-x))
\end{equation}
The diffusivity surrogate function and the predicted data are plotted in Figure \ref{fig:diffusivity}, where the effective vibrational frequency $\nu^*$ is reported to be on the order of $10^{13}$ s$^{-1}$ \cite{VanderVen2000}. The mobility $\widetilde{M}$ is related to $D$ by the equation $\widetilde{M} = x/(k_BT)$ \cite{Jiang2016}. The value of $D$ is multiplied by 4 to approximate the diffusivity at 340 K and is divided by 4 for 260 K. The factor of 4 for every 40 K is based on the predicted diffusivity at 400 K being approximately 30 times the diffusivity at 300 K \cite{VanderVen2000}. Since the free energy for these initial simulations is a function of Li composition only, i.e., a conserved order parameter, the Cahn-Hilliard equation from Eq. (\ref{eqn:CH}) is sufficient to model the system. The non-conservative Allen-Cahn equation in Eq. (\ref{eqn:AC}) is not relevant without inclusion of the non-conserved order parameters, which are obtained from rows 2-7 of the $\widehat{\bsym{Q}}$ matrix presented in Section \ref{sec:order parameters}. The simulations are performed using the finite element method with the \texttt{mechanoChemFEM} code\footnote{Code available at github.com/mechanoChem/mechanoChemFEM}, which is based on the \texttt{deal.II} \cite{dealii2017} library, and run on the ConFlux HPC cluster at the University of Michigan. Adaptive meshing with hanging nodes and adaptive time stepping are used.

\section{Results}
\label{sec:results}

\subsection{Formation energies and configurations from density functional theory}

The predicted average voltage for Li composition ranges [0,1], [0,0.5], and [0.5,1] and increasing values of $U$ are plotted in Figure \ref{fig:volt_U}. By comparing with experimental voltages from Ref. \cite{Aykol2015}, we selected a value of $U = 2.5$ eV to use in the calculation of formation energies for LCO. This provides a good match with the experimental voltage for $x\in[0,1]$ and $x\in[0.5,1]$, with a small error of 0.03 eV for $x\in[0,0.5]$.

\begin{figure}[t]
    \centering
    \includegraphics[width=0.5\textwidth]{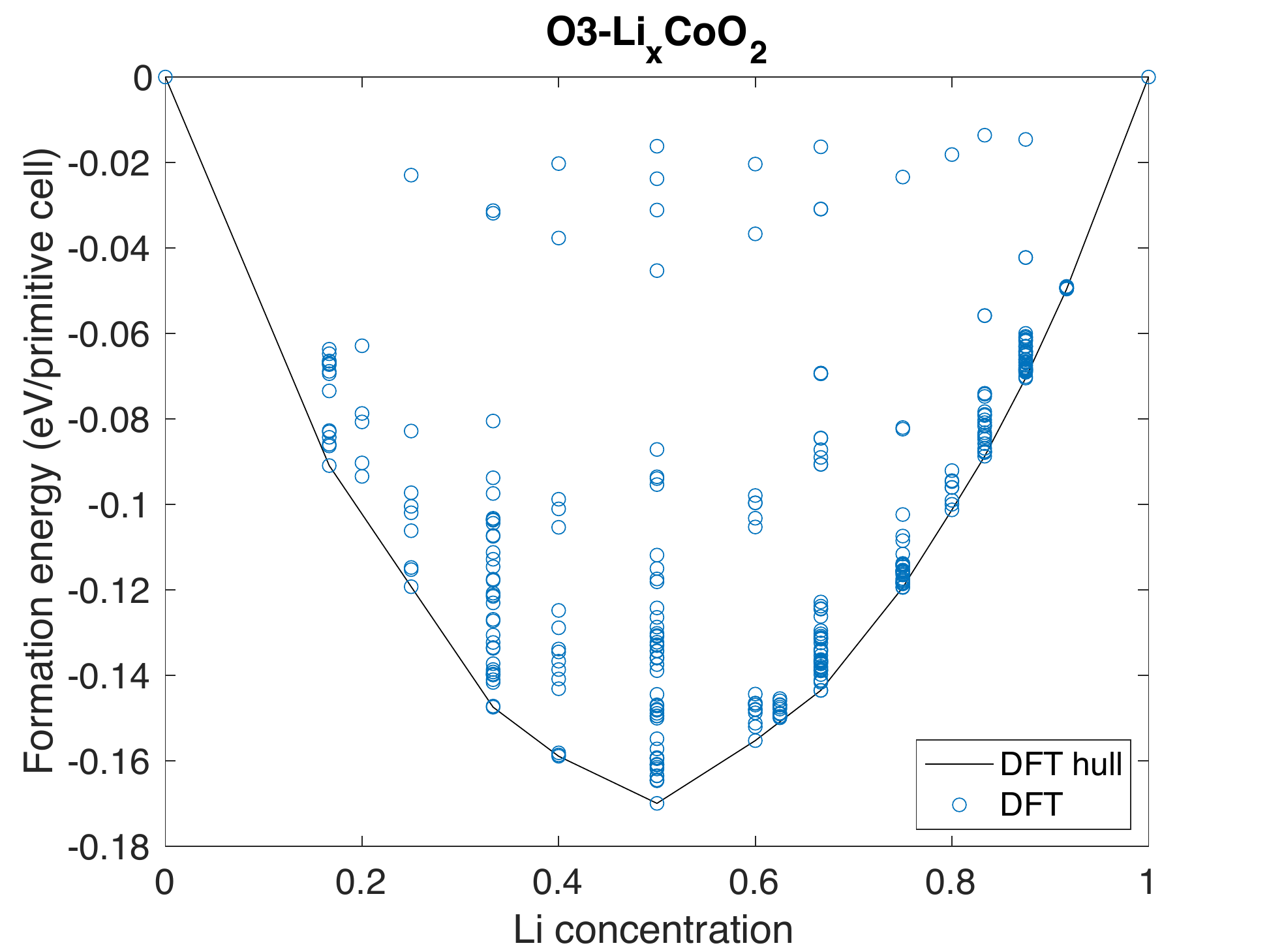}
    \caption{Formation energies and convex hull from the DFT calculations.}
    \label{fig:dft_hull}
\end{figure}

\texttt{CASM} was used to identify a set of configurations with the O3 crystal structure for use in parametrizing the cluster expansion. Using the tuned value for $U$, fully-relaxed DFT calculations were completed for 333 of these configurations, with \texttt{CASM} automatically adjusting the size of the k-points grid according to atomic configuration. Due to difficulty in convergence for some configurations, it was necessary in those cases to incrementally increase $U$ up to 2.5 to achieve a fully converged result. The calculated formation energies and associated convex hull with the predicted ground states are plotted in Figure \ref{fig:dft_hull}. The unit cell volume for LiCoO$_2$ was calculated to be 32.502 \AA$^3$.

\begin{figure}[tb]
        \centering
\begin{minipage}[t]{0.25\textwidth}
        \centering
	\includegraphics[width=0.9\textwidth]{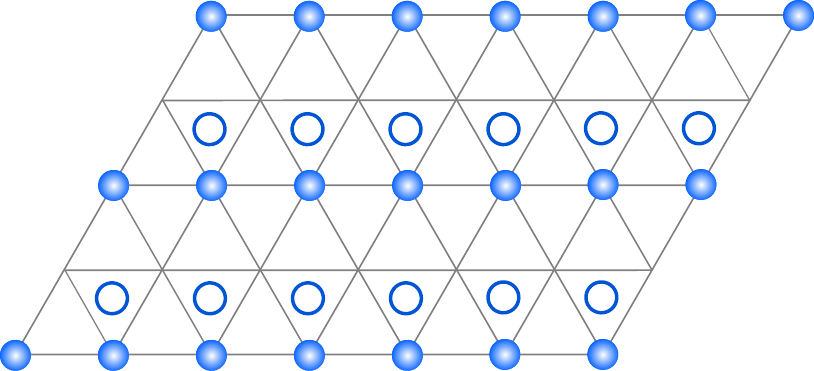}
	\subcaption{Configuration A}
\end{minipage}%
\begin{minipage}[t]{0.25\textwidth}
        \centering
	\includegraphics[width=0.9\textwidth]{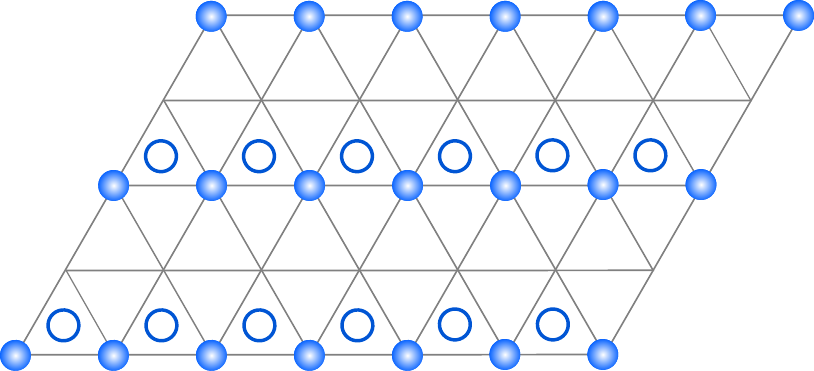}
	\subcaption{Configuration B}
\end{minipage}%
\begin{minipage}[t]{0.25\textwidth}
        \centering
	\includegraphics[width=0.90\textwidth]{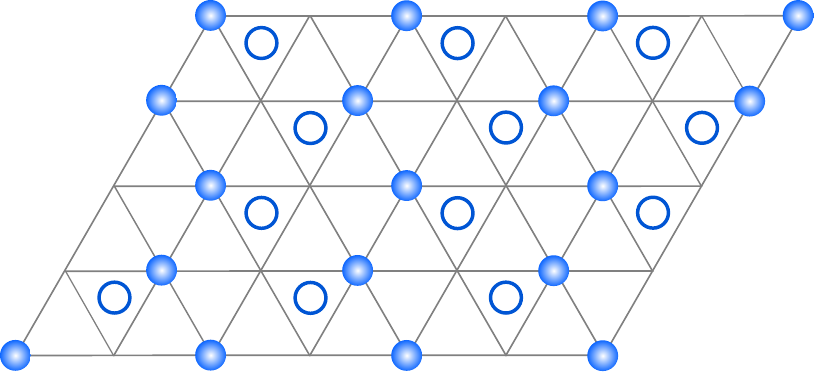}
	\subcaption{Configuration C}
\end{minipage}
\caption{Three configurations with Li concentration $c = 0.5$ with the following calculated formation energies: (a) -0.150, (b) -0.164, and (c) -0.170 eV/primitive cell. The filled circles represent Li atoms within a Li layer, and the empty circles represent Li atoms in the adjacent Li layer.}
\label{fig:c=0.5}
\end{figure}

Three configurations at $x = 0.5$ are shown in Figure \ref{fig:c=0.5} that have been discussed in the literature, all with low formation energies. Note that configurations A and B both exhibit a row ordering and are distinguished only by the relative stacking of the Li layers, whereas configuration C shows a zig-zag ordering. Experimental work has found evidence for the row ordering \cite{Reimers1992,ShaoHorn2003,Takahashi2007}, where Takahashi, et al. \cite{Takahashi2007} present configuration A as the crystal structure and Shao-Horn, et al. \cite{ShaoHorn2003} suggest configuration B might be possible. Charge ordering of Co$^{3+}$ and Co$^{4+}$ atoms has been experimentally observed at low temperatures and Li compositions of $x = 1/2$ and $x = 2/3$ \cite{Motohashi2009}. The development of charge ordering can be monitored through the magnetic moments in the DFT calculations \cite{Aykol2015}. Interestingly, evidence of charge splitting appears in the DFT results in this work for configuration A, but not in configurations B or C. However, a more accurate investigation of charge ordering would require use of advanced DFT functionals such as hybrid exchange functionals~\cite{b3lyp,pbe0,hse} or extended Hubbard models (DFT+$U$+$V$)~\cite{Amaricci2010,campo2010extended}.

In their first-principles study, Van der Ven, et al. \cite{VanderVen1998} found configuration B (as labeled in Figure \ref{fig:c=0.5}) to have the lowest formation energy, followed closely by configuration C. They make the observation, however, that the energy difference is small enough that it is difficult to determine with certainty the true ground state solely from the first-principles calculations. In our work, we take the zig-zag ordering of configuration C to be the ground state based on the calculated formation energies, but we repeat the caveat that its calculated energy is close to that of the configurations with row ordering. The differences in energy between our results and those of earlier work are likely due to our using DFT+$U$, vdW-DF, and generalized gradient approximations (GGA), as opposed to using local-density approximations (LDA).

\subsection{Statistical mechanics}

\subsubsection{Cluster expansion}
\begin{figure}[t]
    \centering
    \includegraphics[width=0.5\textwidth]{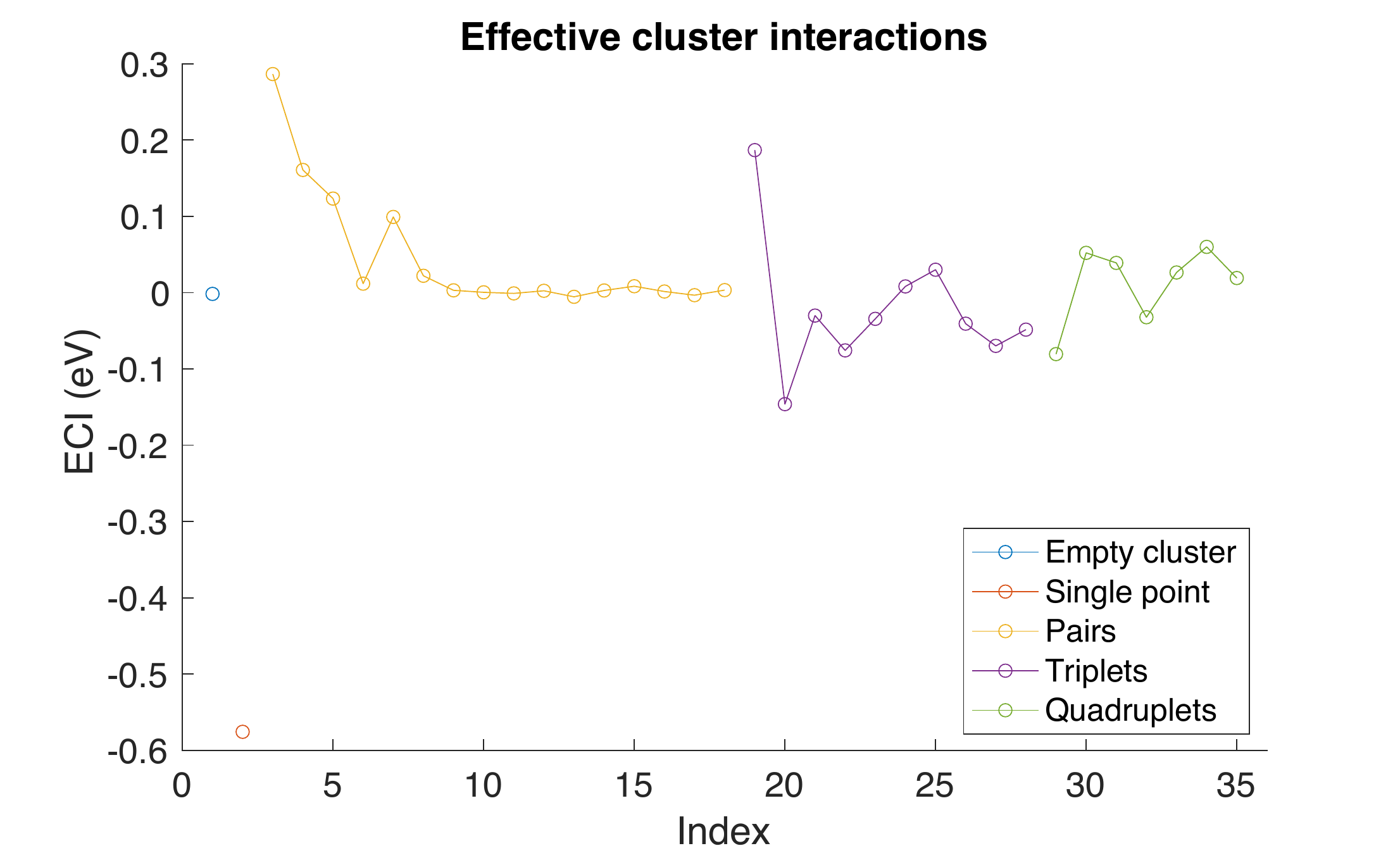}
    \caption{ECI values associated with each basis function in the cluster expansion.}
    \label{fig:ECI}
\end{figure}

\begin{figure}[t]
    \centering
    \includegraphics[width=0.5\textwidth]{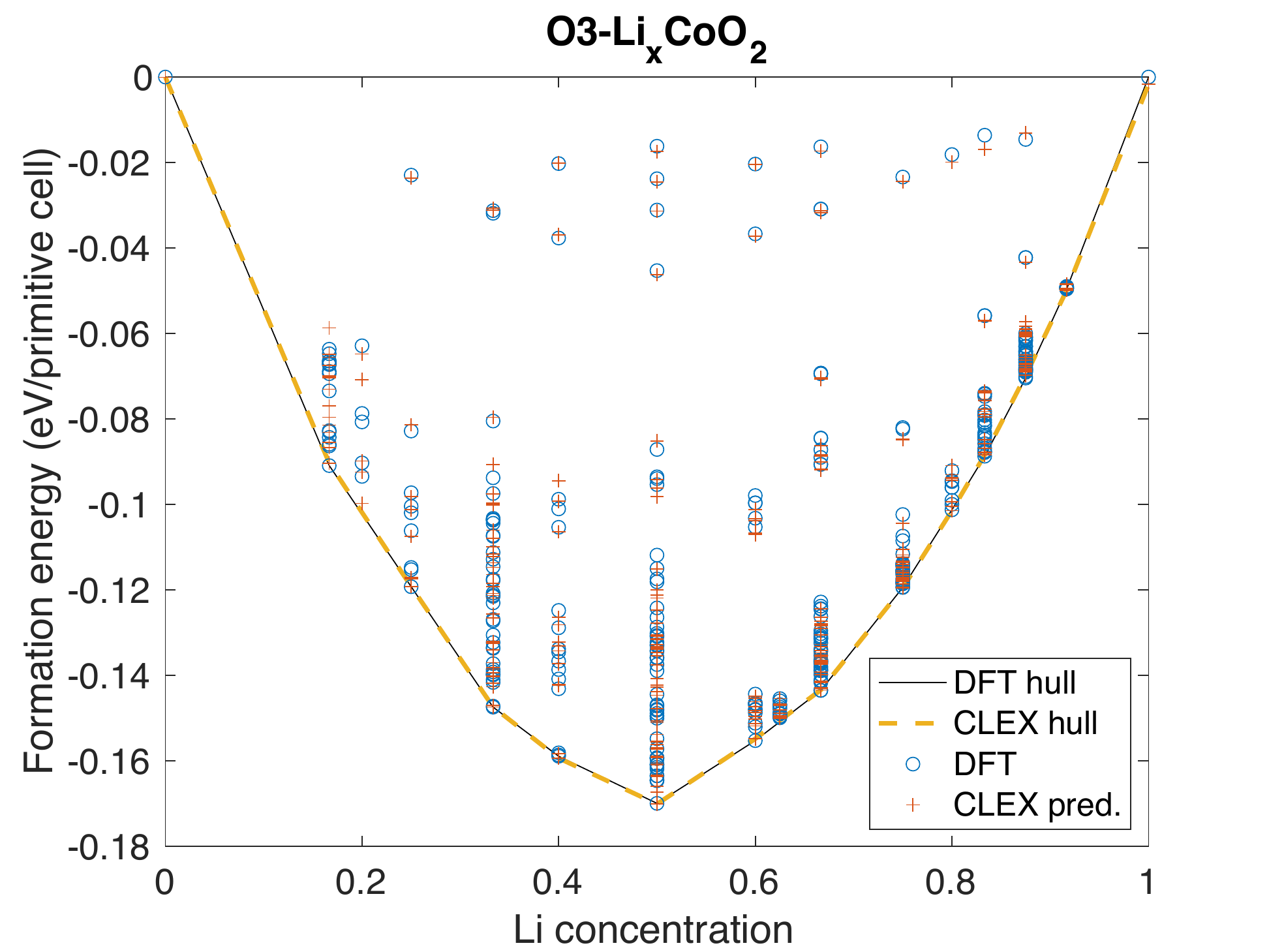}
    \caption{Formation energies and convex hull predicted by the cluster expansion (CLEX) are compared with those from the DFT calculations. The configurations predicted by the cluster expansion to lie on the convex hull are the same as those on the DFT convex hull.}
    \label{fig:hull}
\end{figure}

In parameterizing the cluster expansion, the genetic algorithm selected 35 basis functions. The RMS error for the cluster expansion prediction of the formation energy was 2.49 meV per primitive cell. A plot of the effective cluster interaction (ECI) for each cluster in the cluster expansion is shown in Figure \ref{fig:ECI}. The formation energies predicted by the cluster expansion are also plotted with the DFT calculated formation energies in Figure \ref{fig:hull}, along with the their respective convex hulls. It can be seen the convex hull constructed from the cluster expansion predictions matches the convex hull of the DFT data.

\subsubsection{Chemical potential data and phase diagram from Monte Carlo simulations}

\begin{figure}[t]
    \centering
    \includegraphics[width=0.4\textwidth]{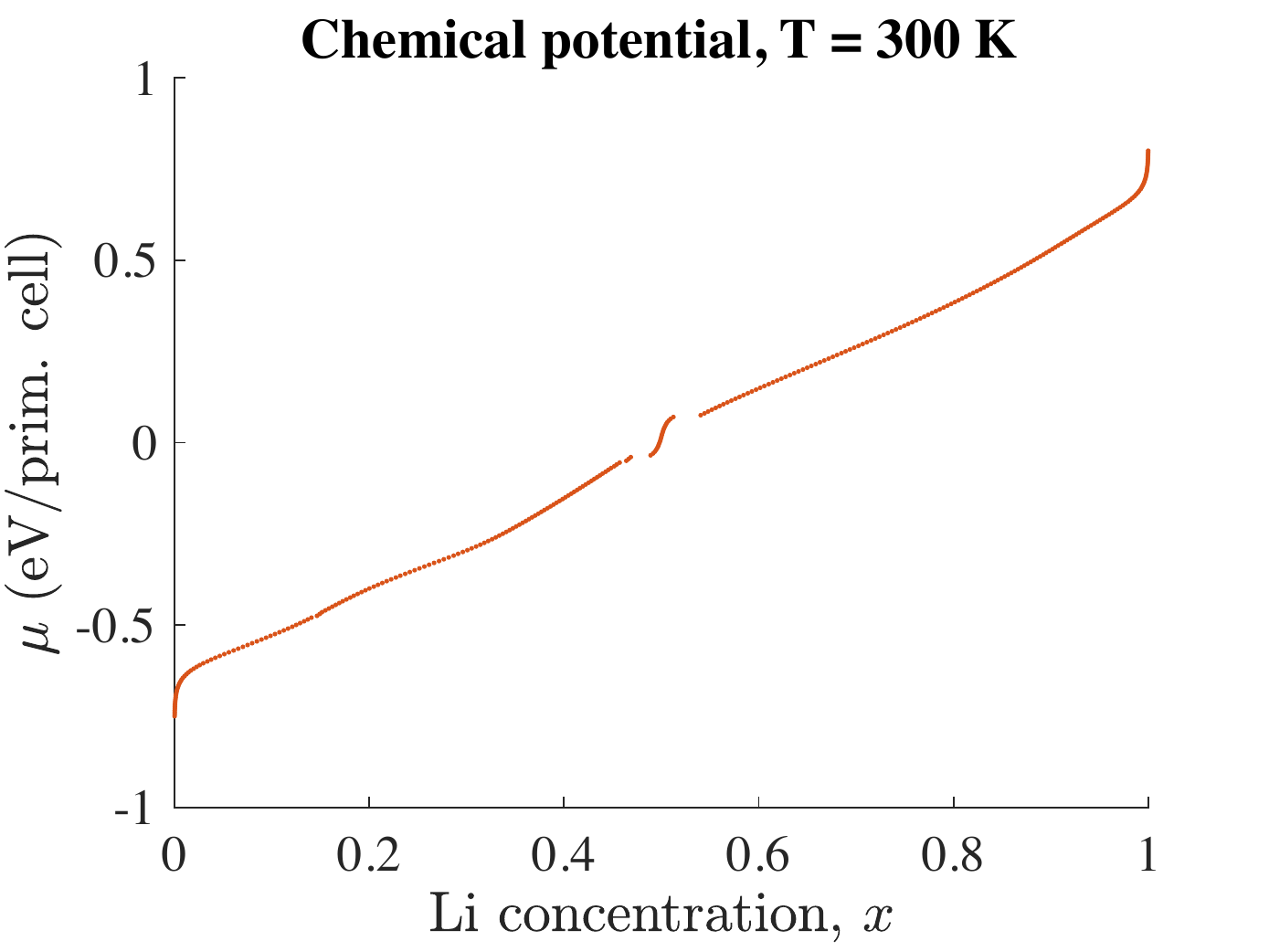}
    \caption{Chemical potential values at 300 K, predicted using Monte Carlo simulations without bias potentials.}
    \label{fig:mu_300K}
\end{figure}

\begin{figure}[t]
    \centering
    \includegraphics[width=0.5\textwidth]{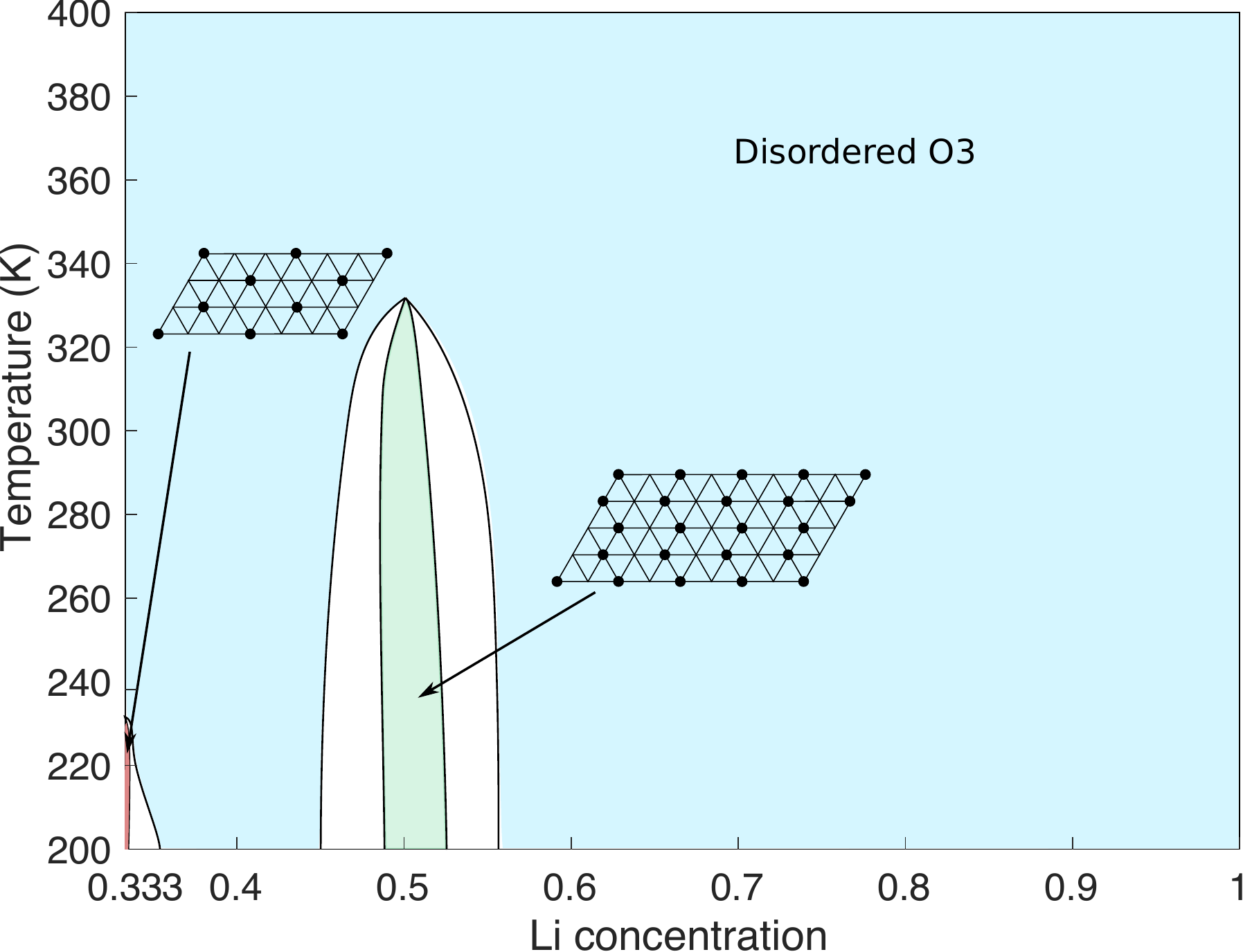}
    \caption{Phase diagram for the O3 structure for LCO with orderings at $x=1/3$ and $x=1/2$, based on Monte Carlo results.}
    \label{fig:phase_diagram}
\end{figure}

Li content as a function of chemical potential and temperature was computed through Monte Carlo statistical mechanics calculations for temperatures ranging from 200 K to 400 K, initially without using the bias potentials. Since the bias potentials were not used in these calculations the two-phase regions are easily seen as gaps in the composition values of the data (see, for example, Figure \ref{fig:mu_300K}).
Regions with orderings are recognized by the steepness and positive slope of the curve traced out by connecting the chemical potential data points. This indicates a strongly convex free energy well and thermodynamic preference for these structures over the corresponding composition intervals.
Two-phase regions appear on either side of these intervals. These data were used to construct an approximate phase diagram for O3 LCO, for Li compositions greater than $x \ge 1/3$. The calculated phase diagram of the Li$_x$CoO$_2$ system is shown in Figure \ref{fig:phase_diagram}.

\begin{figure}[t]
    \centering
    \includegraphics[width=0.45\textwidth]{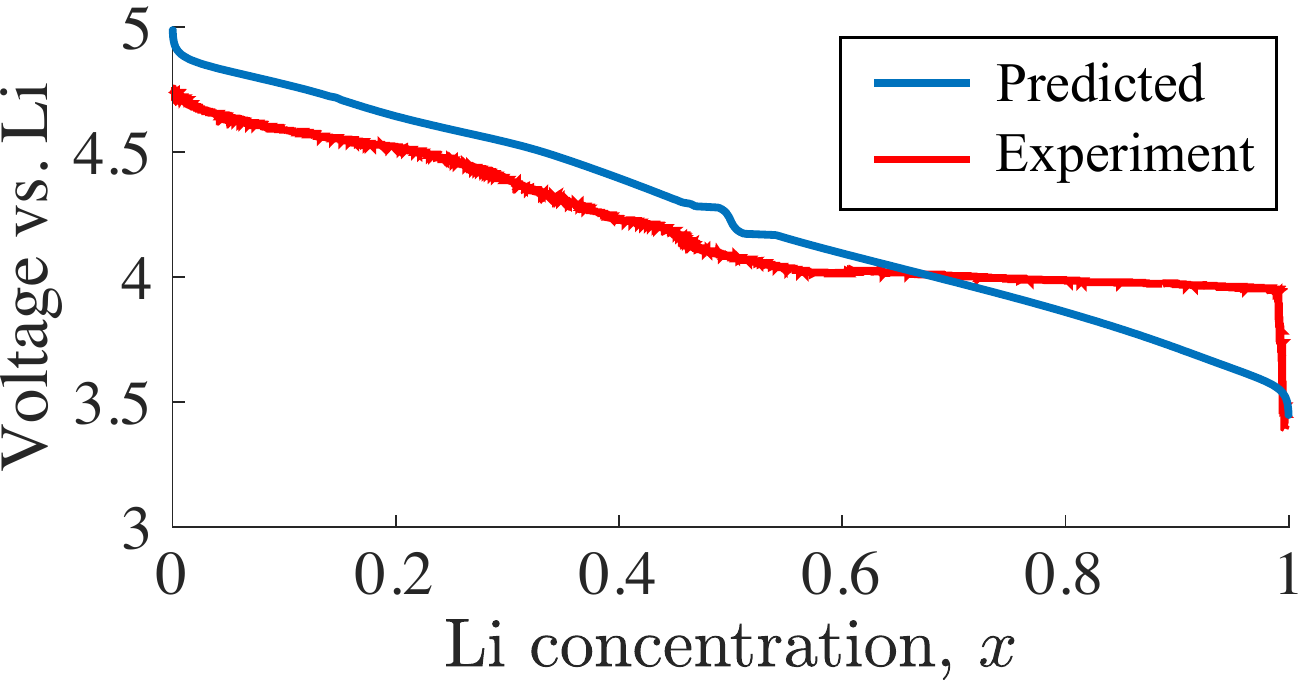}
    \caption{Comparison of the predicted voltage at 300 K (blue) with the experimental voltage (red) from Ref. \cite{Amatucci1996}.}
    \label{fig:voltage}
\end{figure}

From the Monte Carlo results, the order-disorder transition temperature for the ordering at $x=0.5$ is slightly greater than 330 K, which is consistent with experimental results \cite{Reimers1992}. An ordering is also predicted by the Monte Carlo results at $x = 0.33$, appearing below approximately 235 K. This ordering has not been seen experimentally at room temperature, but some reported evidence shows that it may exist at low temperatures \cite{ShaoHorn2003}, which is consistent with our calculations. The improvement in accuracy of the predicted order-disorder temperatures over previous first-principles work by Van der Ven, et al. \cite{VanderVen1998} is likely due to advancements in the DFT methods over the last two decades. A fairly good agreement between our computed voltage profile at 300 K and the experimental measurements \cite{Amatucci1996} is also observed in Figure \ref{fig:voltage}. Similar to past first-principles studies on LCO, however, because of the lack of the metal-insulator transition that leads to the two-phase region observed between $\sim 0.7$-$0.9$ in calculations, the voltage plateau in the same region is not captured. We also note that the predicted voltage, as with the other Monte Carlo results in this work, is only for the O3 structure and, therefore, only valid for compositions greater than $x\approx1/3$.

\begin{figure}[tb]
        \centering
\begin{minipage}[t]{0.28\textwidth}
        \centering
	\includegraphics[width=0.99\textwidth]{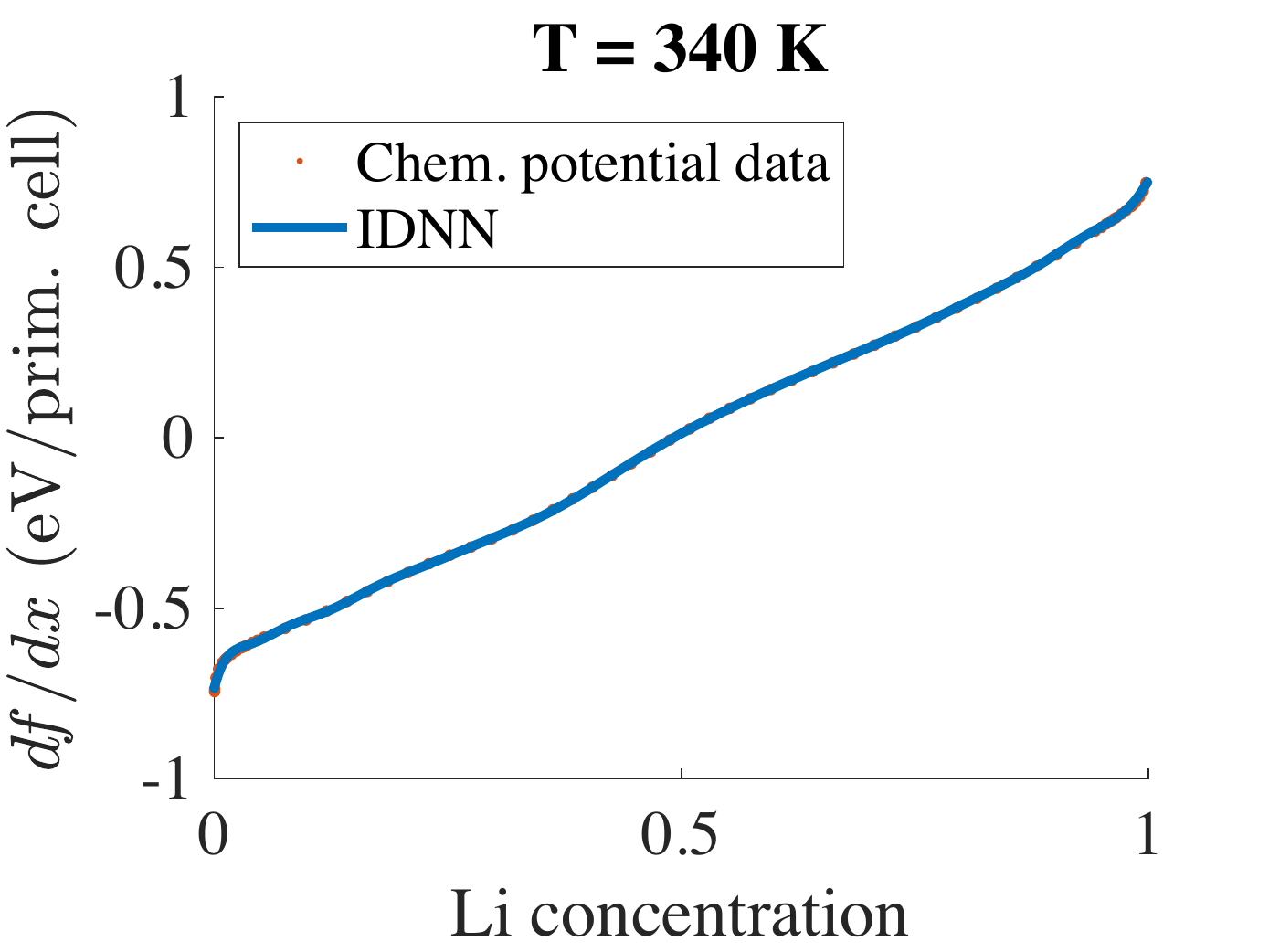}
\end{minipage}%
\begin{minipage}[t]{0.28\textwidth}
        \centering
	\includegraphics[width=0.99\textwidth]{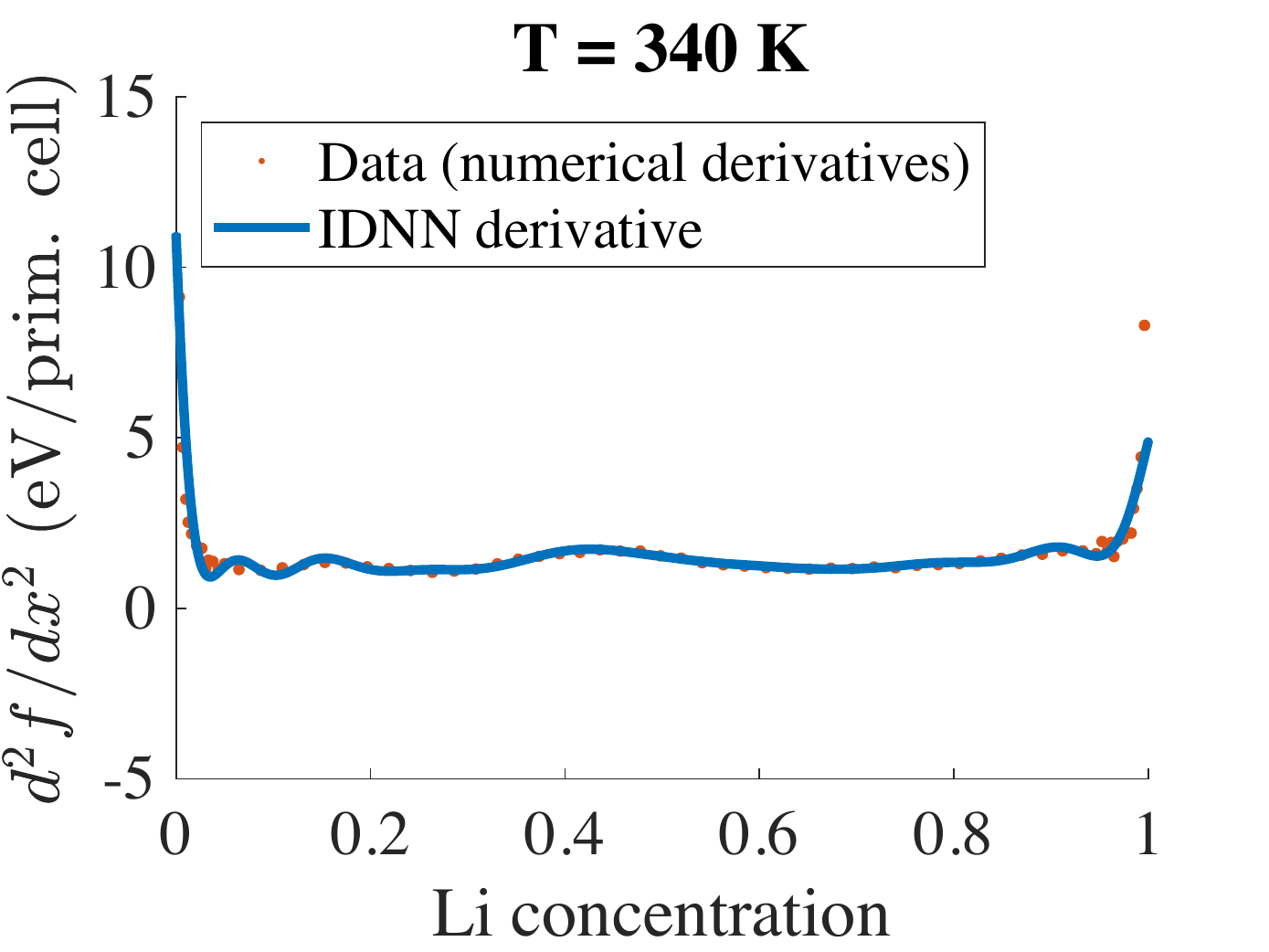}
\end{minipage}%
\begin{minipage}[t]{0.28\textwidth}
        \centering
	\includegraphics[width=0.99\textwidth]{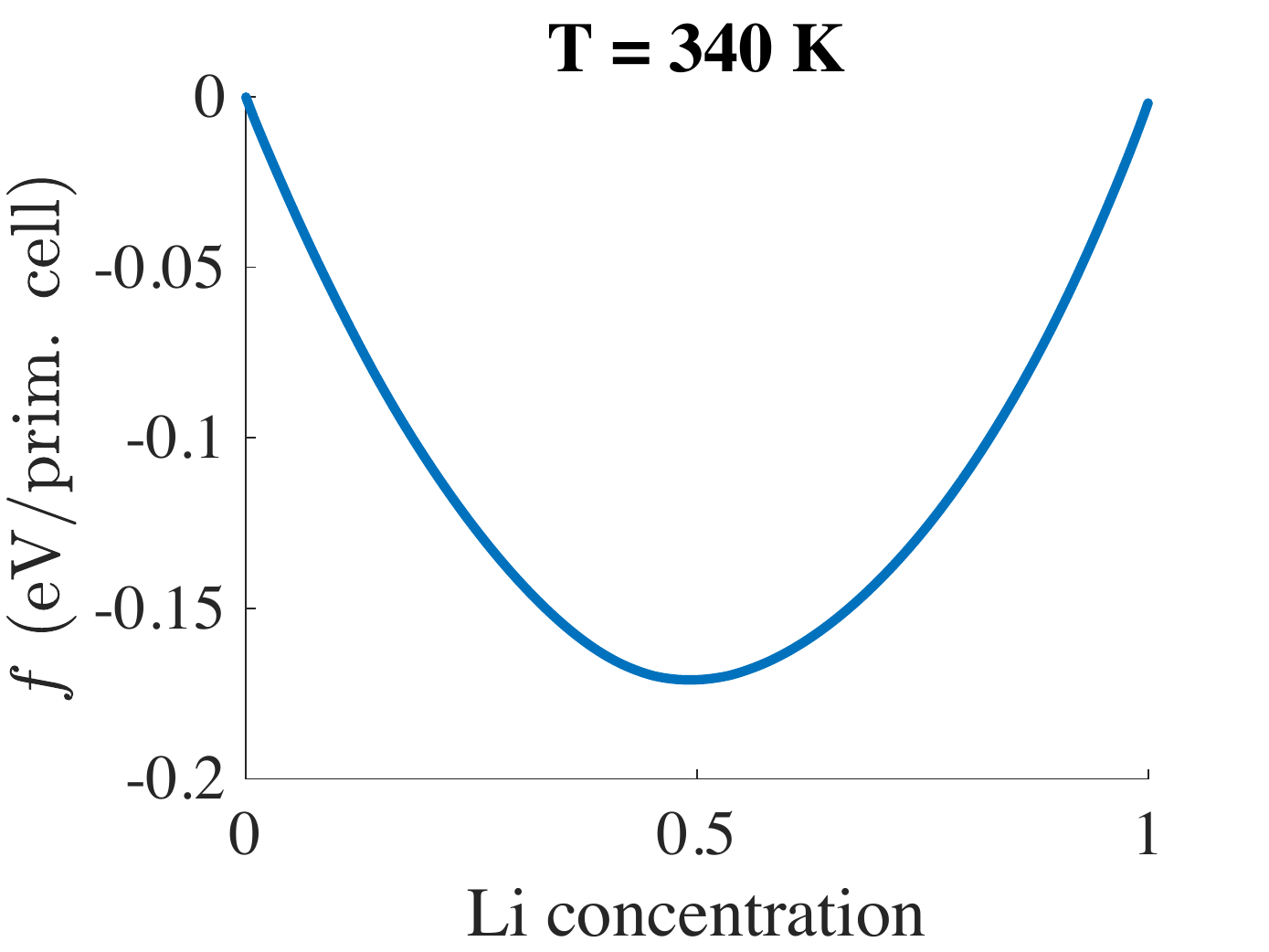}
\end{minipage}
\begin{minipage}[t]{0.28\textwidth}
        \centering
	\includegraphics[width=0.99\textwidth]{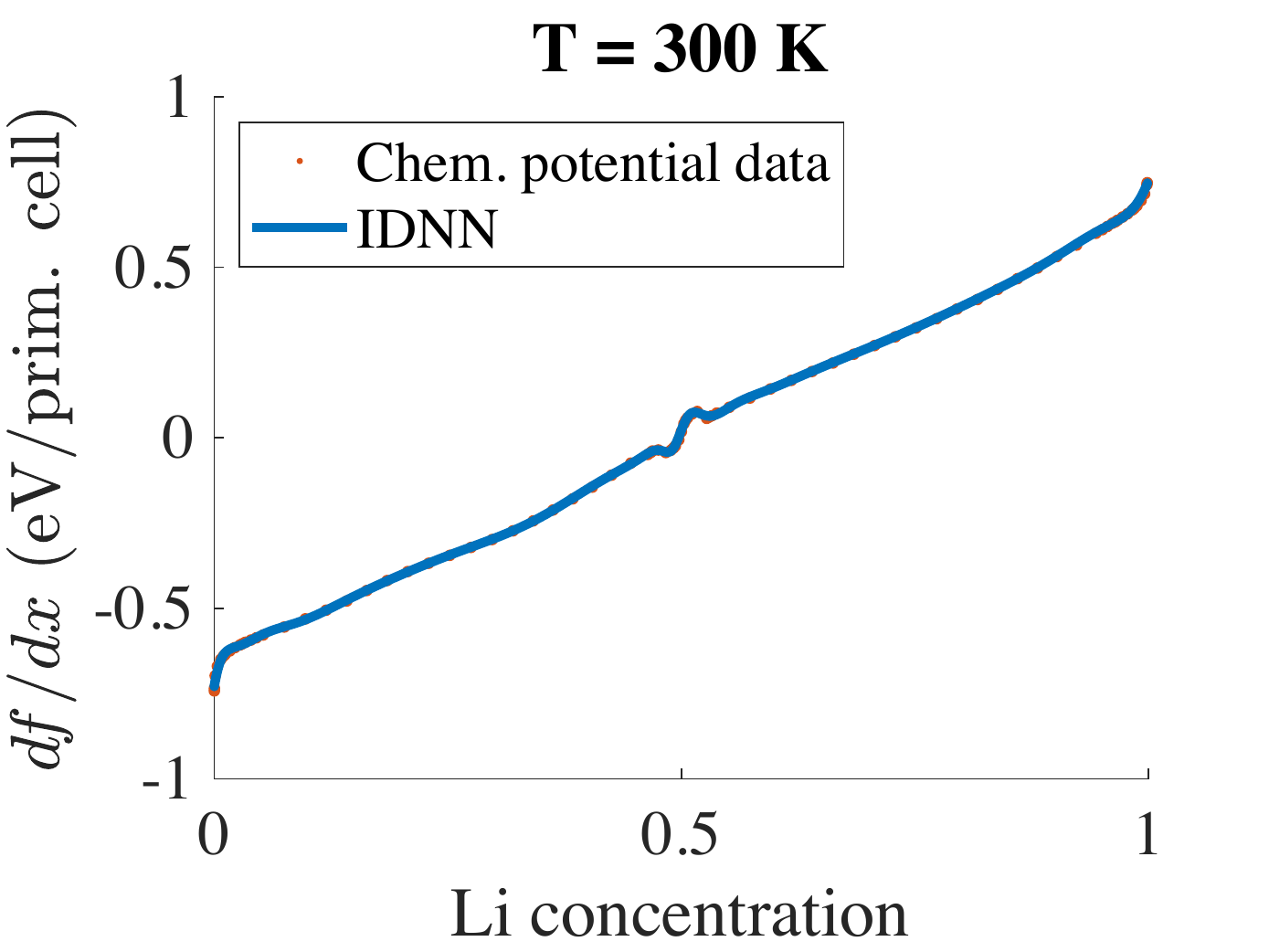}
\end{minipage}%
\begin{minipage}[t]{0.28\textwidth}
        \centering
	\includegraphics[width=0.99\textwidth]{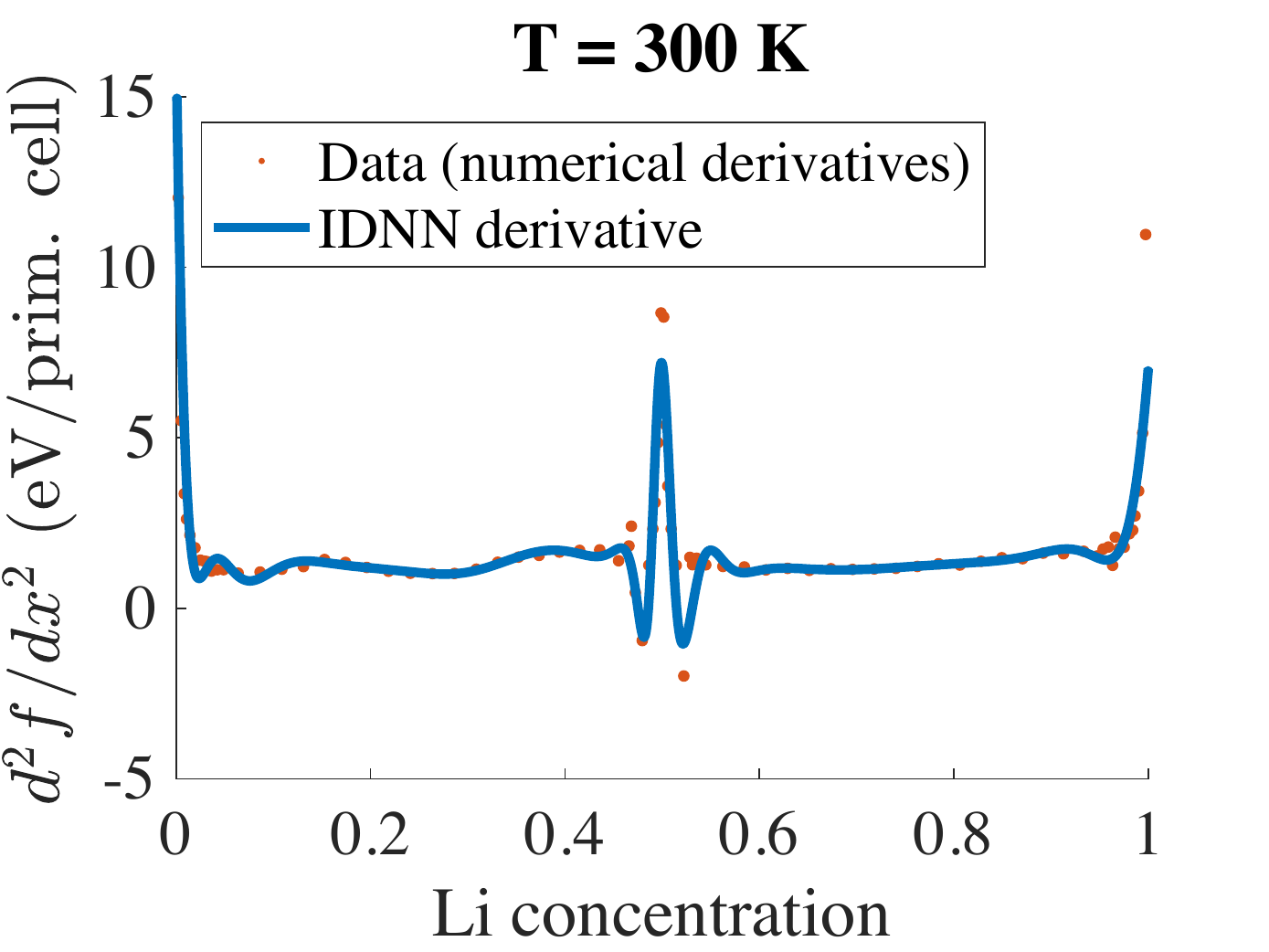}
\end{minipage}%
\begin{minipage}[t]{0.28\textwidth}
        \centering
	\includegraphics[width=0.99\textwidth]{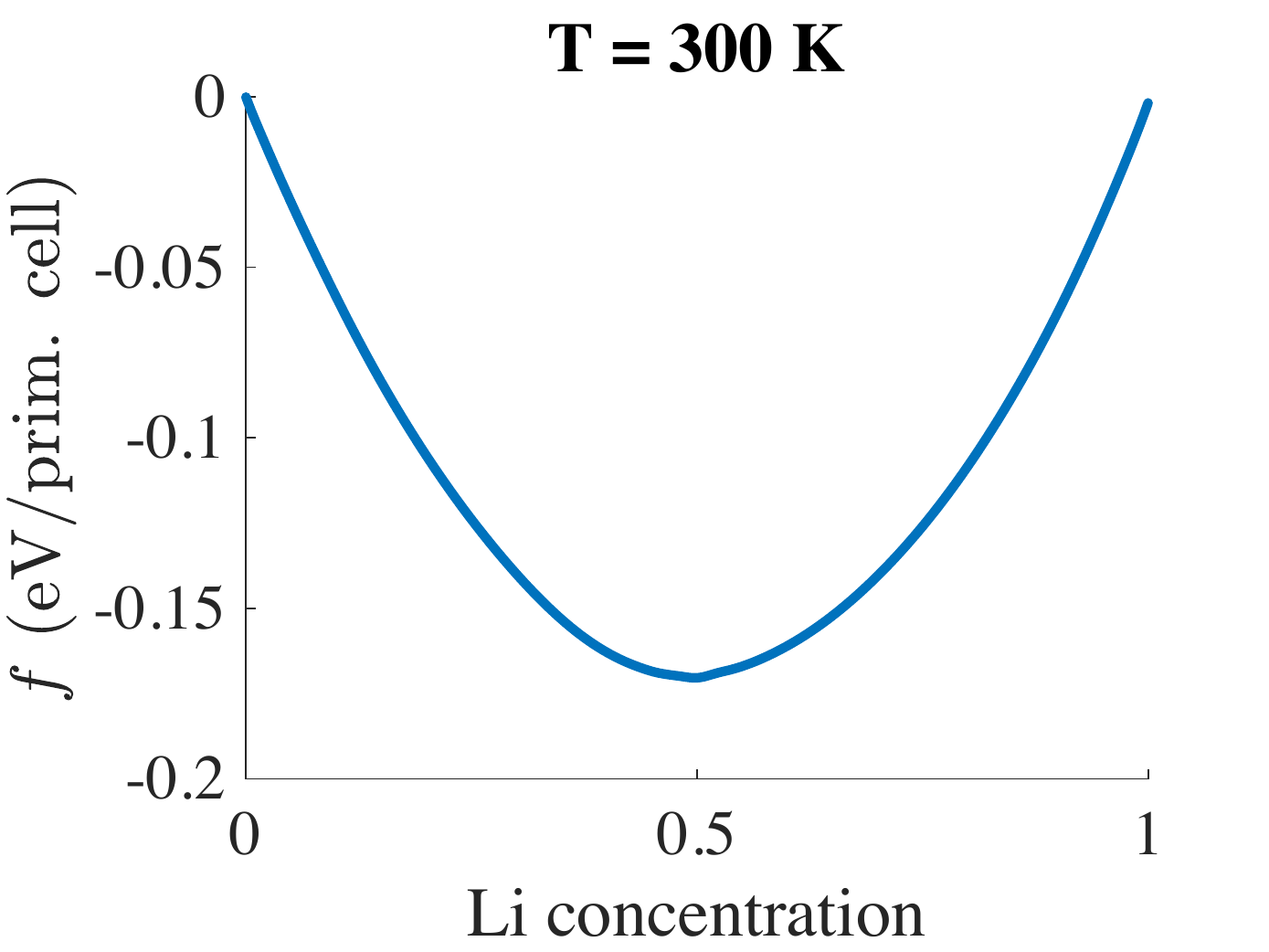}
\end{minipage}
\begin{minipage}[t]{0.28\textwidth}
        \centering
	\includegraphics[width=0.99\textwidth]{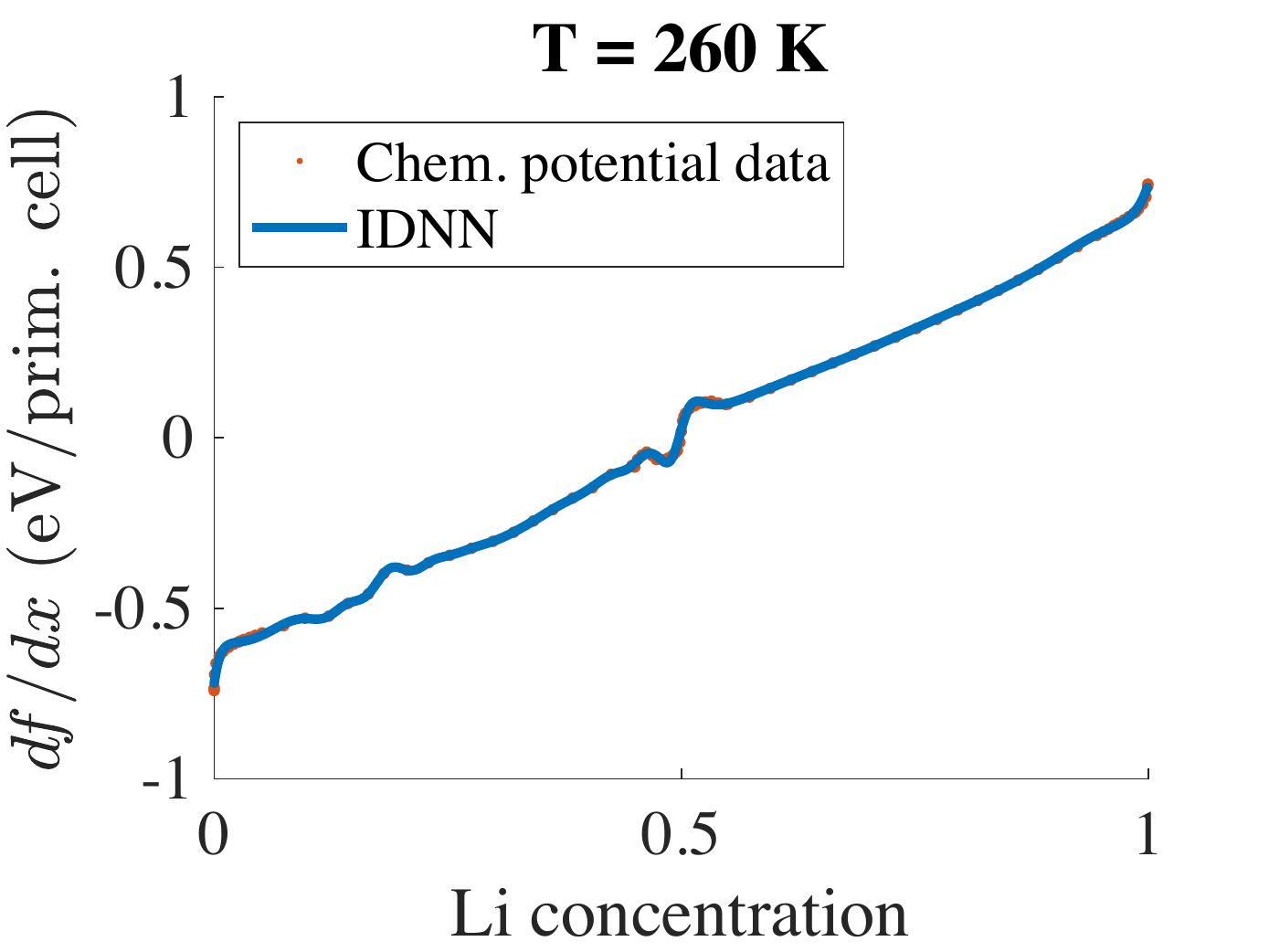}
	\subcaption{Chemical potential, $\mathrm{d}f/\mathrm{d}c$}
\end{minipage}%
\begin{minipage}[t]{0.28\textwidth}
        \centering
	\includegraphics[width=0.99\textwidth]{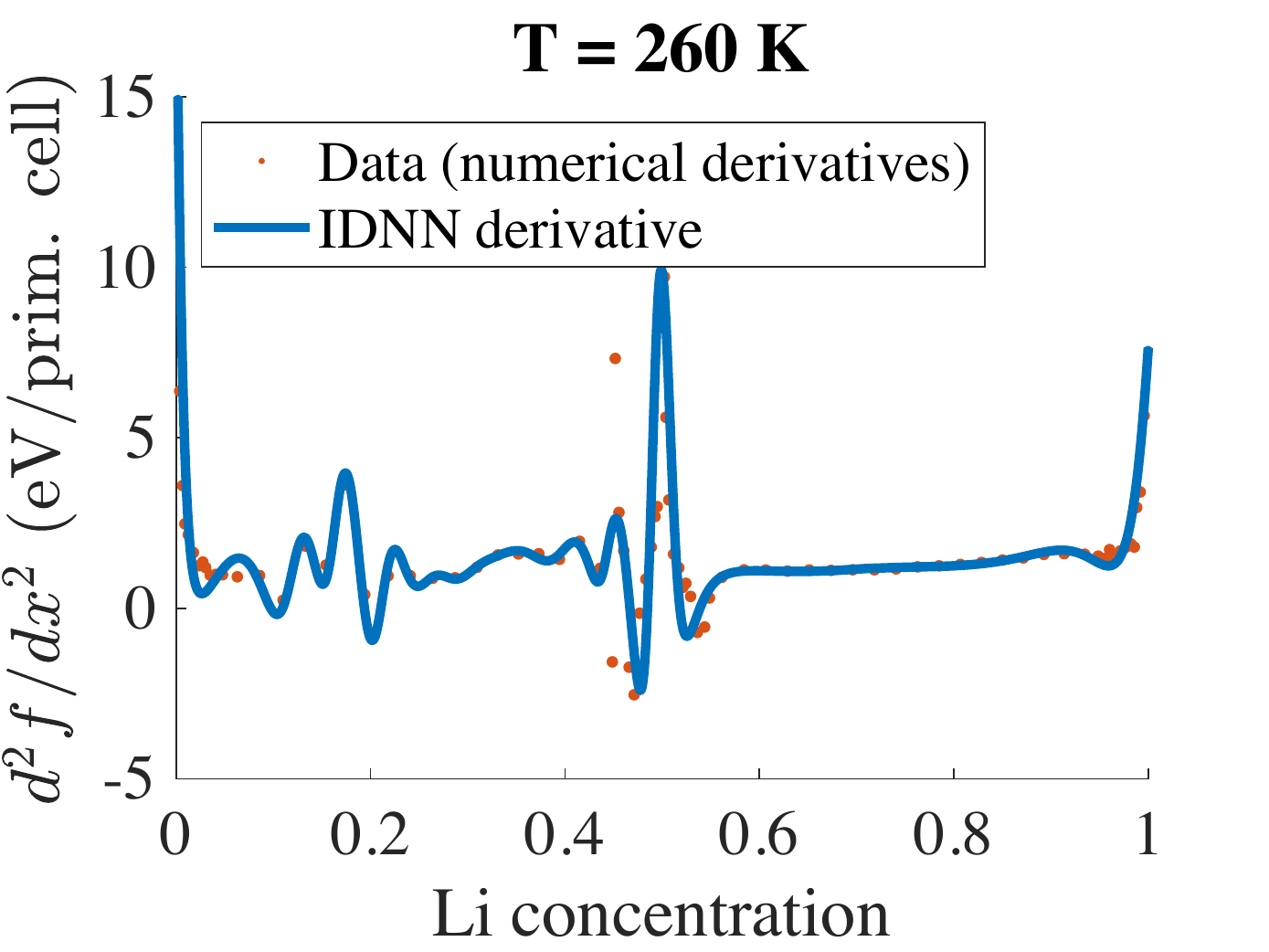}
	\subcaption{$\mathrm{d}^2f/\mathrm{d}c^2$}
\end{minipage}%
\begin{minipage}[t]{0.28\textwidth}
        \centering
	\includegraphics[width=0.99\textwidth]{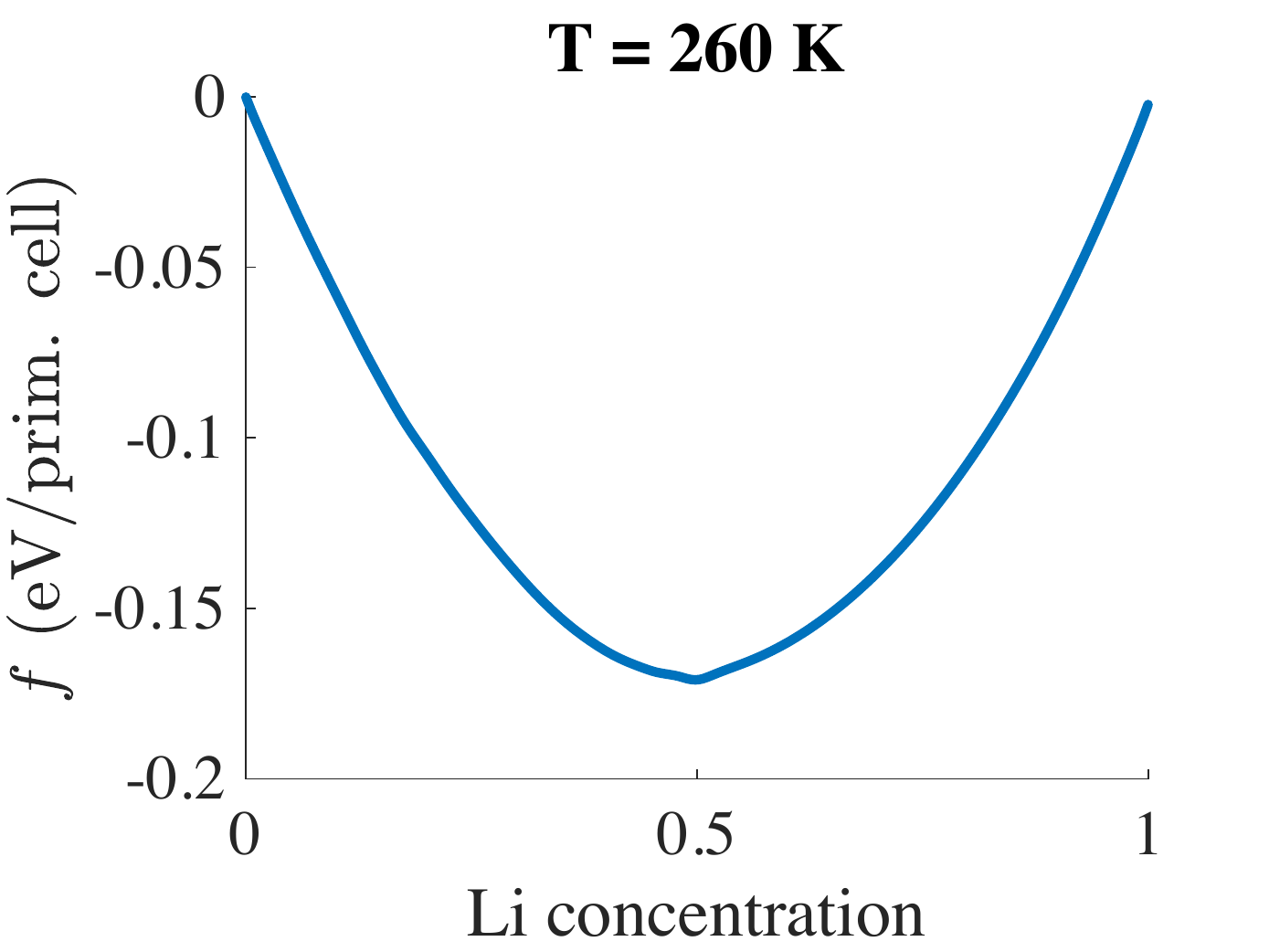}
	\subcaption{Free energy, $f(c)$}
\end{minipage}
\caption{Comparison of IDNN fits with the data for 260 K, 300 K, and 340 K: (a) the chemical potential data used for training, as sampled with the bias potentials, and the IDNN prediction, (b) numerically differentiated data and derivative of the chemical potential IDNN, and (c) the analytically integrated free energy DNN.}
\label{fig:idnn_fit}
\end{figure}

Additional Monte Carlo calculations were performed for temperatures of 260 K, 300 K, and 340 K, this time using bias potentials to sample within the unstable regions of order-disorder transition. Since the fluctuations in the chemical potential around $x=0.5$ are critical to representing the ordering in the phase field models at temperatures below $\sim$330 K, dense sampling of chemical potential data was performed in that region for 260 K and 300 K. Additional sampling was also performed near the extreme compositions of $x=0$ and $x=1$ to help capture the divergent nature of the chemical potential (see Figure \ref{fig:idnn_fit}(a)).

\subsubsection{Symmetry-adapted order parameters}
\label{sec:order parameters}
\begin{figure}[t]
    \centering
    \includegraphics[width=0.8\textwidth]{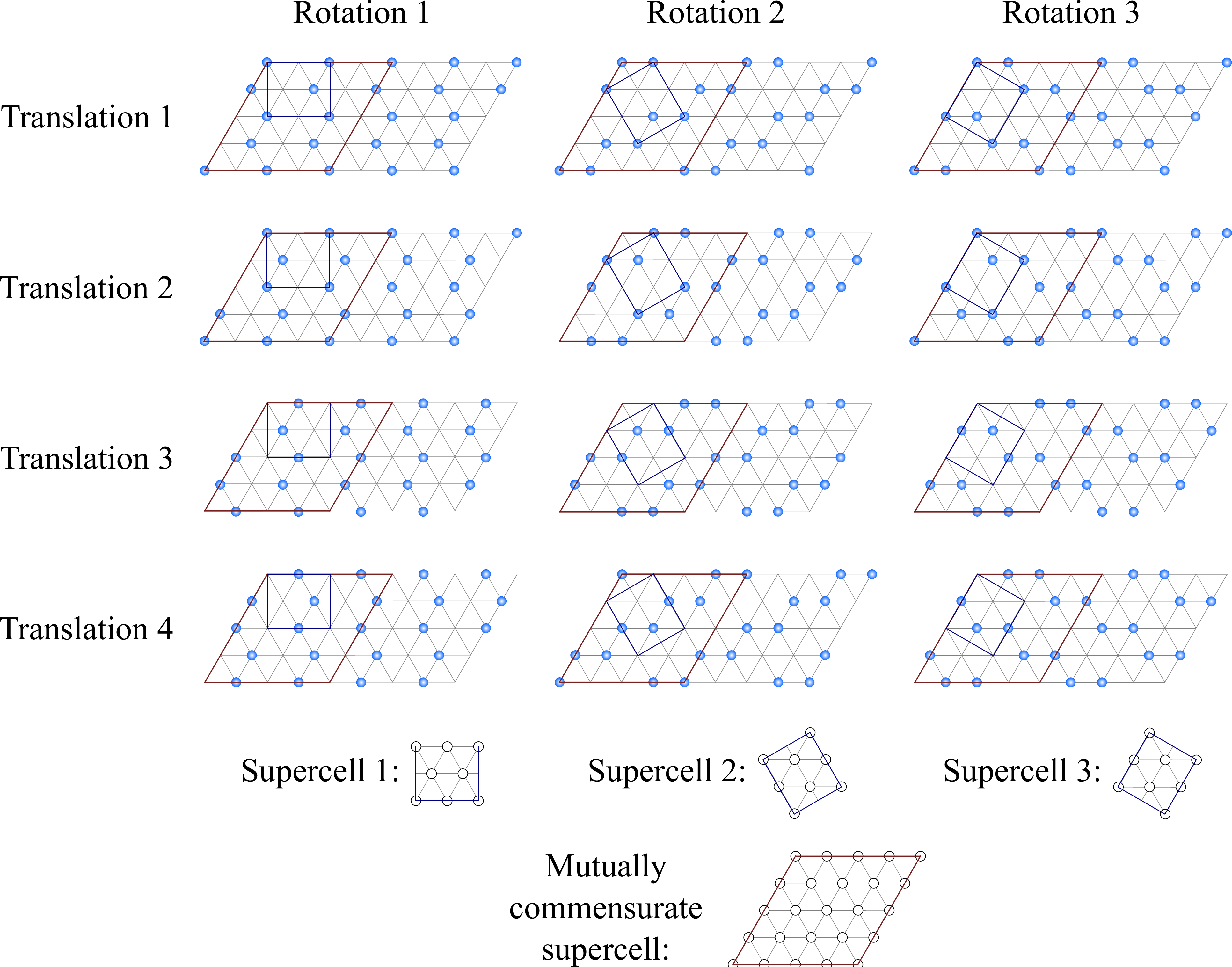}
    \caption{The zig-zag ordering has 12 variants, resulting from combinations of 3 rotations and 4 translations.}
    \label{fig:orderparams}
\end{figure}

Symmetry-adapted order parameters are useful for tracking the order-disorder transition and the different ordering variants that can form. Since the Monte Carlo and DFT results show a zig-zag ordering forming at $x = 0.5$, we developed the corresponding order parameters here. The zig-zag ordering of Li atoms on the triangular lattice has 12 variants (3 rotations $\times$ 4 translations), as shown in Figure \ref{fig:orderparams}. Following the algorithm to construct the mutually commensurate supercell \cite{Natarajan2017} yielded a supercell with 16 unique sublattice sites. Using the 16 sublattice sites as a basis, the twelve variants can be written as a set of twelve vectors $\{\bsym{x}^{(1)},\ldots,\bsym{x}^{(12)}\}$.

Additionally, 48 symmetry operations between these twelve variants were enumerated in the group $P$, each represented by a $16 \times 16$ matrix. Following the algorithm from Thomas and Van der Ven \cite{thomas2017}, we constructed a $P$-invariant matrix and performed its eigenvalue decomposition. It resulted in four nonzero eigenvalues (one distinct and three repeated) and four sets of eigenvectors, with a total of 16 eigenvectors. The eigenvectors formed the rows of a $16 \times 16$ transformation matrix $\bsym{Q}$, to map a sublattice composition vector $\bsym{x}$ to a vector $\bsym{\eta}$ with 16 order parameters.

The subset of order parameters relevant to a given ordering can be identified by operating with $\bsym{Q}$ on the vectors $\bsym{x}^{(i)}$ describing the variants of that ordering. The superscript in $\bsym{x}^{(i)}$ indexes the variants. Components of $\bsym{\eta}$ that are zero for all variants are irrelevant for describing that ordering, and the corresponding rows can be removed from $\bsym{Q}$. Multiplying $\bsym{Q}$ by the vectors describing the twelve zig-zag variants, it was determined that seven order parameters were relevant. One was associated with the distinct eigenvalue and corresponded to the composition averaged over all 16 sublattice sites. The other six order parameters were associated with one of the degenerate eigenvalues which had six corresponding eigenvectors. Thus, a reduced $7 \times 16$ matrix $\widehat{\bsym{Q}}$ can be used if only the zig-zag ordering is of interest, where the first row gives the average composition, such that $\eta_0 = x$:
\setcounter{MaxMatrixCols}{16}
\begin{align}
    \widehat{\bsym{Q}} &= \frac{1}{16}
    \setlength\arraycolsep{1pt}
    \begin{bmatrix*}[r]
    1 &  1 &  1 &  1 &  1 &  1 &  1 &  1 &  1 &  1 &  1 &  1 &  1 &  1 &  1 &  1\\
    1 & -1 &  1 & -1 & -1 &  1 & -1 &  1 & -1 &  1 & -1 &  1 &  1 & -1 &  1 & -1\\
    1 &  1 & -1 & -1 & -1 &  1 &  1 & -1 & -1 & -1 &  1 &  1 &  1 & -1 & -1 &  1\\
  -1 &  1 & -1 &  1 & -1 &  1 & -1 &  1 &  1 & -1 &  1 & -1 &  1 & -1 &  1 & -1\\
    1 & -1 & -1 &  1 & -1 &  1 &  1 & -1 &  1 & -1 & -1 &  1 & -1 &  1 &  1 & -1\\
    1 & -1 & -1 &  1 &  1 &  1 & -1 & -1 & -1 &  1 &  1 & -1 & -1 & -1 &  1 &  1\\
    1 &  1 & -1 & -1 & -1 & -1 &  1 &  1 &  1 &  1 & -1 & -1 & -1 & -1 &  1 &  1
    \end{bmatrix*}
\end{align}
Of the other two degenerate eigenvalues, one had three associated eigenvectors and described the row ordering, and the other had six eigenvectors and described a double-row ordering.

\subsection{Trained integrable deep neural networks}

The chemical potential IDNNs for 260 K, 300 K, and 340 K, trained as functions of Li composition, are plotted on top of the chemical potential data in Figure \ref{fig:idnn_fit}(a), showing a fairly accurate match.
Since the regions of phase separation occur where the second derivative of the free energy is negative, the chemical potential data were numerically differentiated and plotted with the analytical derivatives of the chemical potential IDNNs, as shown in Figure \ref{fig:idnn_fit}(b).
Phase separation regions do indeed appear in both the Monte Carlo data, where they are resolved with the biased potential, and in the neural network fit on either side of the ordering at $x=0.5$ at 260 K and 300 K, although the two-phase region at $x\approx 0.52$ for 260 K appears to occur at a slightly lower composition in the IDNN than in the data.
The analytically integrated free energy neural networks are plotted in Figure \ref{fig:idnn_fit}(c).
We also emphasize that these data and functions represent the O3 structure and are, therefore, most relevant for $x \ge 1/3$, since previous experimental and computational work show that LCO adopts the H1-3 structure below $x = 1/3$.

\subsection{Phase field simulations}
\label{sec:phasefieldresults}

Here, we demonstrate our integrated scale bridging framework via three case studies of phase field simulations that help gain some insight to the phase stability of LCO cathodes in the composition neighborhood of the order-disorder transition around $x = 0.5$. The thermodynamic description entering the phase field model is quantitatively accurate, since it is the result of the scale bridging framework. Furthermore, we use appropriate kinetic data. However, some physics remains to be included in our model, such as strain effects, which are known to cause LCO particle degradation \cite{wang1999tem,choi2006particle,pender2020electrode}, interphase interface energies, explicit representations of the different orderings, and spatio-temporal heat generation/transport. Therefore, we do not claim to present fully quantitative predictions, but demonstrate at least how morphology evolves under various microstructure-specific, environmental, kinetic and cycling conditions at a resolution not accessible in experiments and at a computational fidelity not previously accessed with regard to the length scales. From such simulations we make connections to observations in the experimental literature on degradation of LCO \cite{wang1999tem,choi2006particle,pender2020electrode,leng2015effect}.

\begin{figure}[tb]
    \centering
    \includegraphics[width=0.9\textwidth]{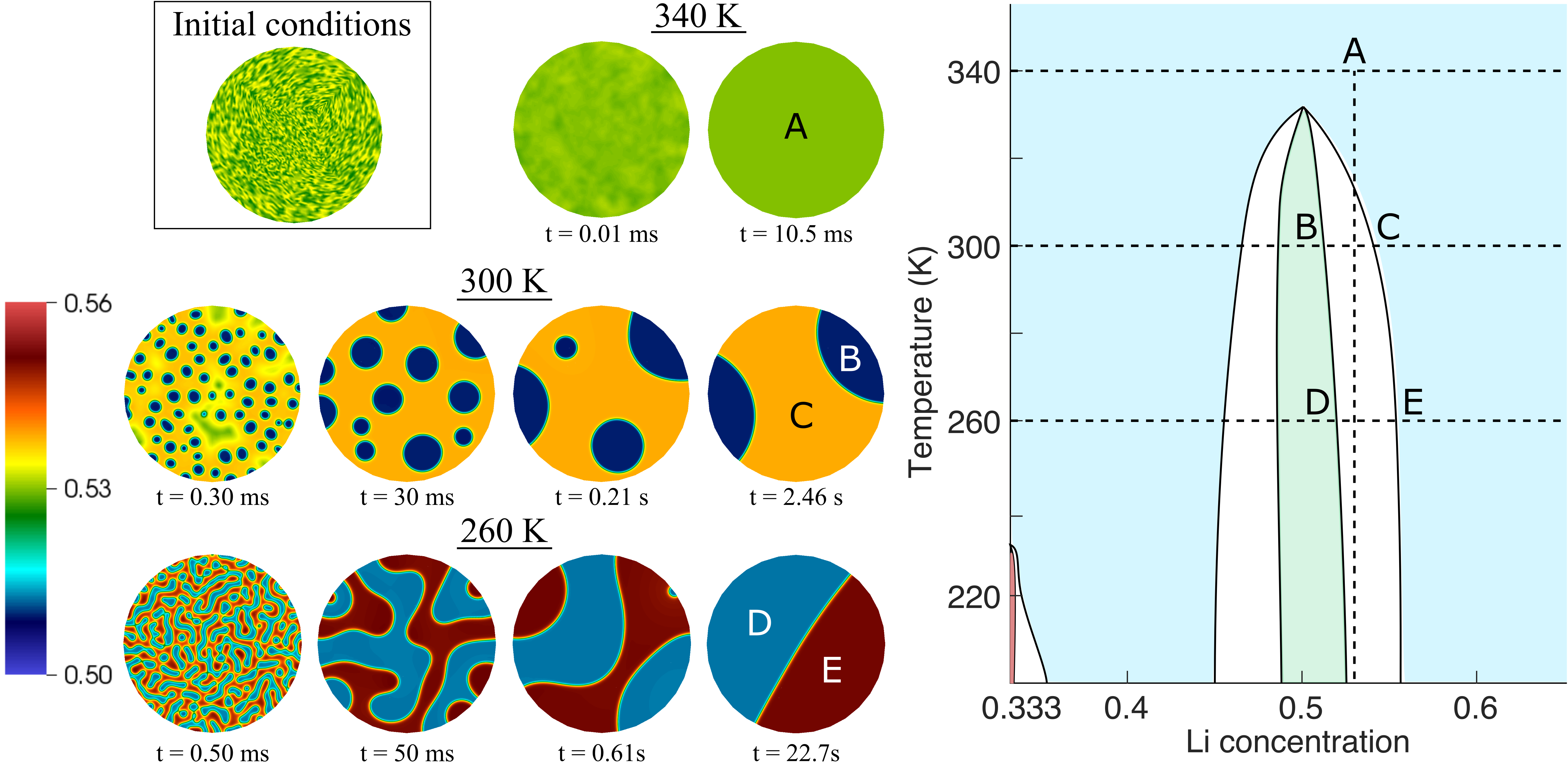}
\caption{2D phase field simulation results at 260, 300, and 340 K showing the Li composition in a 1 $\mu$m diameter particle, with initial Li composition randomly perturbed about $x = 0.53$ and no boundary flux. The zoomed-in phase diagram on the right identifies the phases that are observed in the simulations.}
\label{fig:spinodal}
\end{figure}

In the first set of phase field simulations, the Li composition was initialized as a randomly perturbed field about $x=0.53$ for each of the three temperatures, within a circular domain with a diameter of 1 $\mu$m. No fluxes were applied, and the composition field was allowed to evolve until a steady state was achieved. The simulation at 340 K quickly reached a uniform composition of $x=0.53$, consistent with the completely convex free energy at that temperature. At 300 K and 260 K, however, spinodal decomposition took place, followed by Ostwald ripening, as shown in Figure \ref{fig:spinodal}. In both simulations, there was separation into an ordered phase and a disordered phase. Slight differences in the compositions of these phases can be seen, however, at the different temperatures. At 260 K, the ordered phase appears at $x = 0.512$ and the disordered phase at $x = 0.551$, whereas the ordered and disordered phases occur at $x = 0.509$ and $x = 0.539$, respectively, at 300 K. This shift toward higher compositions at the lower temperature for this particular two-phase region is qualitatively consistent with the phase diagram in Figure \ref{fig:phase_diagram}, although slight mismatches between the nonconvex region of the IDNN and the data, particularly at 260 K, has shifted the predicted compositions slightly lower than the values shown in the phase diagram. The fact that the phase boundaries occur at different compositions for different temperatures is also seen in the ratio of ordered to disordered phase in the transient microstructures observed at 300 and 260 K. Since the ratio of ordered to disordered phase is smaller at 300 K, circular precipitates (in 2D) of the ordered phase appear, whereas the ordered phase regions are elongated and more tortuous at 260 K.

\begin{figure}[tb]
    \centering
\begin{minipage}[t]{0.7\textwidth}
        \centering
	\includegraphics[width=0.99\textwidth]{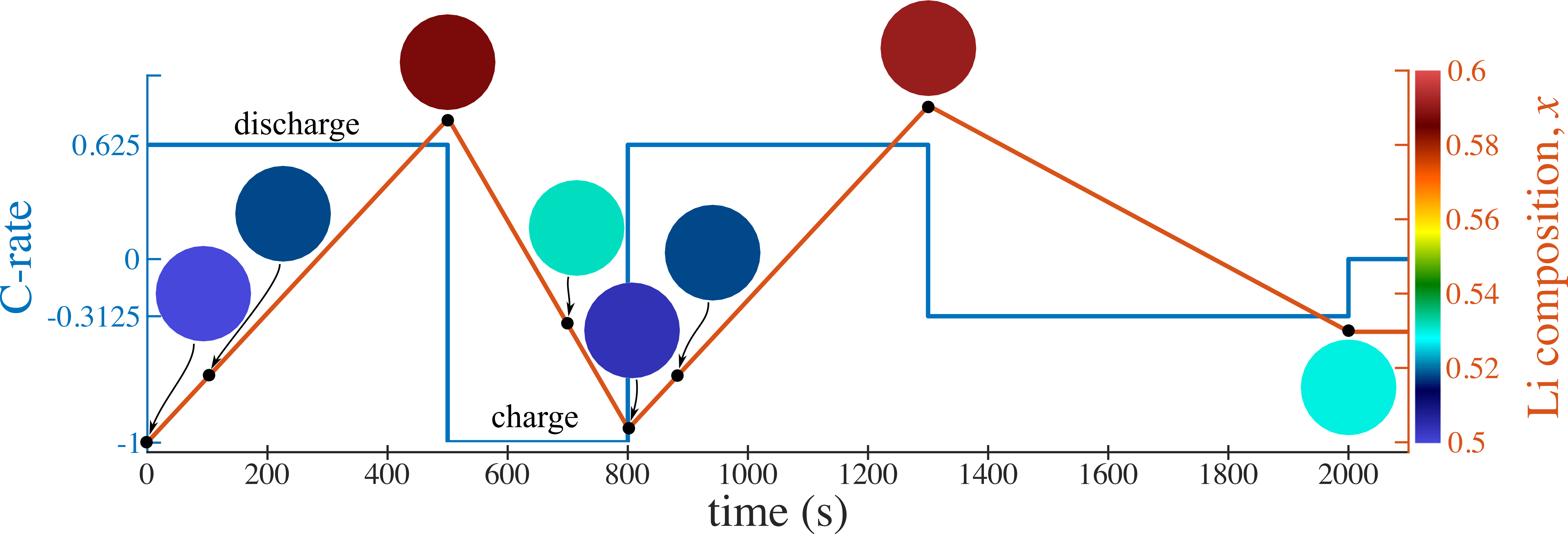}
	\subcaption{340 K}
\end{minipage}
\begin{minipage}[t]{0.7\textwidth}
        \centering
	\includegraphics[width=0.99\textwidth]{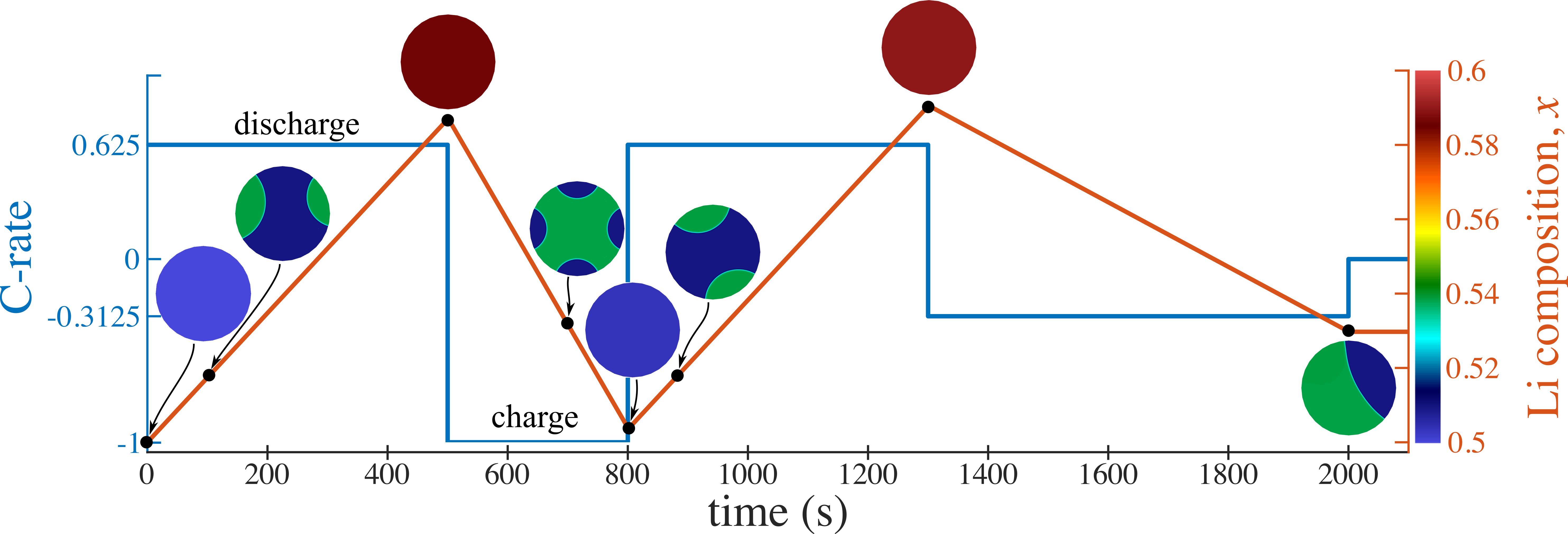}
	\subcaption{300 K}
\end{minipage}
\begin{minipage}[t]{0.7\textwidth}
        \centering
	\includegraphics[width=0.99\textwidth]{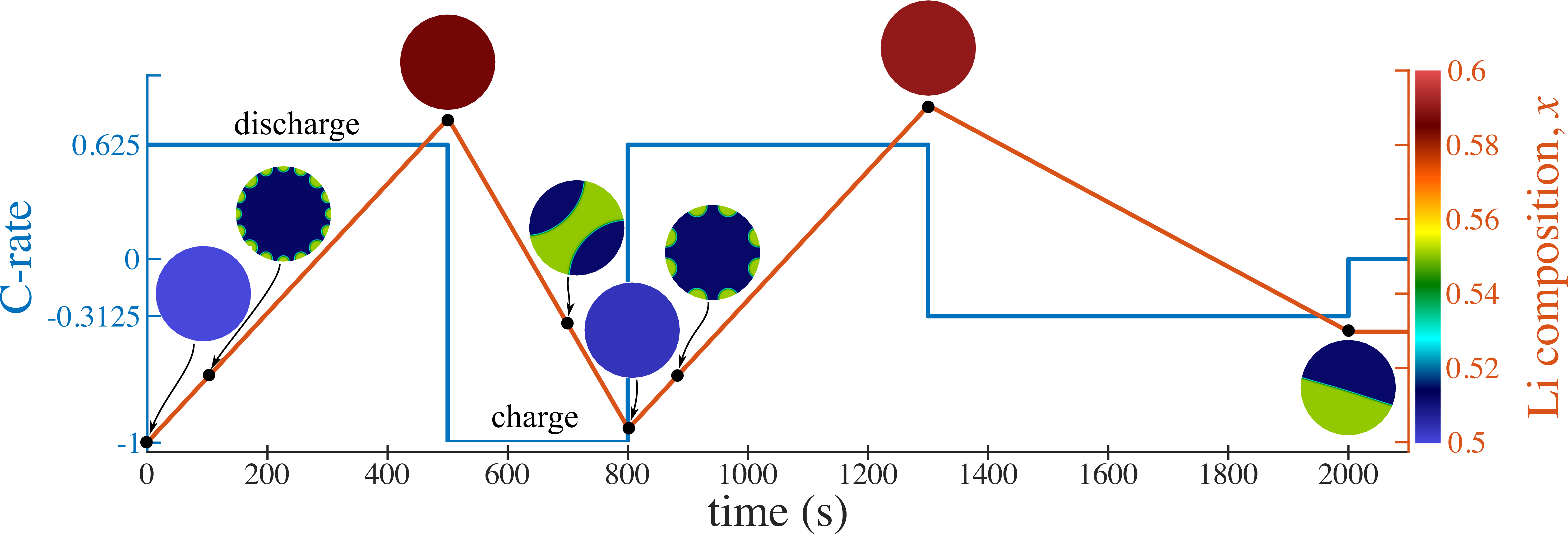}
	\subcaption{260 K}
\end{minipage}
\caption{Phase field results showing the Li composition field resulting from applying a cycling current density for (a) 340, (b) 300, and (c) 260 K, for a 1 $\mu$m diameter particle. The applied C-rate is set to 0.625, -1, 0.625, -0.3125, and 0 C at 0, 500, 800, 1300, and 2000 s, respectively, where a positive sign denotes discharging and negative is charging. The blue line shows the applied C-rate, and the red line shows the corresponding average Li composition of the particle.}
\label{fig:cycle}
\end{figure}

As seen in Figure \ref{fig:spinodal}, notable variations can arise in the morphology of the charged cathode particle in a temperature range that overlaps the window of relevance for actual batteries. For example, based on the figure, battery designers should anticipate morphology changes in charged lithium ion batteries in an electric vehicle parked outdoors in the summer during day \emph{versus} night in many geographical locations. As observed by Pender et al. \cite{pender2020electrode}, lattice dimension changes in the neighborhood of $x = 0.5$ cause mechanical damage. Order-disorder transitions are pronounced over the $260-300$ K range in Figure \ref{fig:spinodal} and as the associated lattice dimension vary, the mismatches across the order-disorder interphase interfaces will cause stresses that may play a role in driving the damage \cite{wang1999tem}. We also note that the higher temperature disordered phase has a more homogeneous microstructure, and the simulation in Figure \ref{fig:spinodal} predicts that it persists across the composition range studied. This could motivate doping with elements that stabilize the disordered structure. Such a mechanism has relevance to the Al and La doped stabilization of the order-disorder transition as observed recently \cite{pender2020electrode}. The phase field treatment also predicts higher driving forces for Li flux in regions of rapidly varying interface curvature, which are seen at the lower temperatures. The lower mobility at $260$ K relative to $300$ K slows down transport despite the correspondingly greater curvature-induced driving force.

Additional phase field simulations show the effect of cycling of a Li$_x$CoO$_2$ particle at 260, 300, and 340 K, as presented in Figure \ref{fig:cycle} \cite{Jiang2016}. The Li composition was initialized to 0.5, following the recommended lower limit of composition for LCO, below which large c-axis variations of the lattice cause enhanced degradation \cite{pender2020electrode}. This corresponds to an initial predicted voltage of 4.23 V. An initial current density of 0.1335 A m$^{-2}$ was applied. The applied current density was adjusted to -0.2137, 0.1335, -0.0668, and 0 A m$^{-2}$ at 500, 800, 1300, and 2000 s, respectively. This cycling corresponds to C-rates of 0.625, -1, 0.625, -0.3125, and 0 C, where a positive sign denotes discharging and negative is charging. This cycle results in a net discharging, as indicated by the voltage-composition plot in Figure \ref{fig:voltage}, with a final predicted voltage of 4.14, 4.17, or 4.19 V for 260, 300, or 340 K, respectively. Equilibrium was reached in all simulations by 2013 s. Since 340 K is above the order-disorder transition at $x=0.5$, no ordered phase forms. However, the presence of the order-disorder interfaces at lower temperatures suggests that such regimes of charging/discharging of LCO particles can induce transient misfit stresses. Such cycling between the ordered and disordered states at the lower temperatures can promote degradation as discussed in the experimental literature \cite{choi2006particle,leng2015effect,pender2020electrode}.

\begin{figure}[tb]
    \centering
    \includegraphics[width=0.6\textwidth]{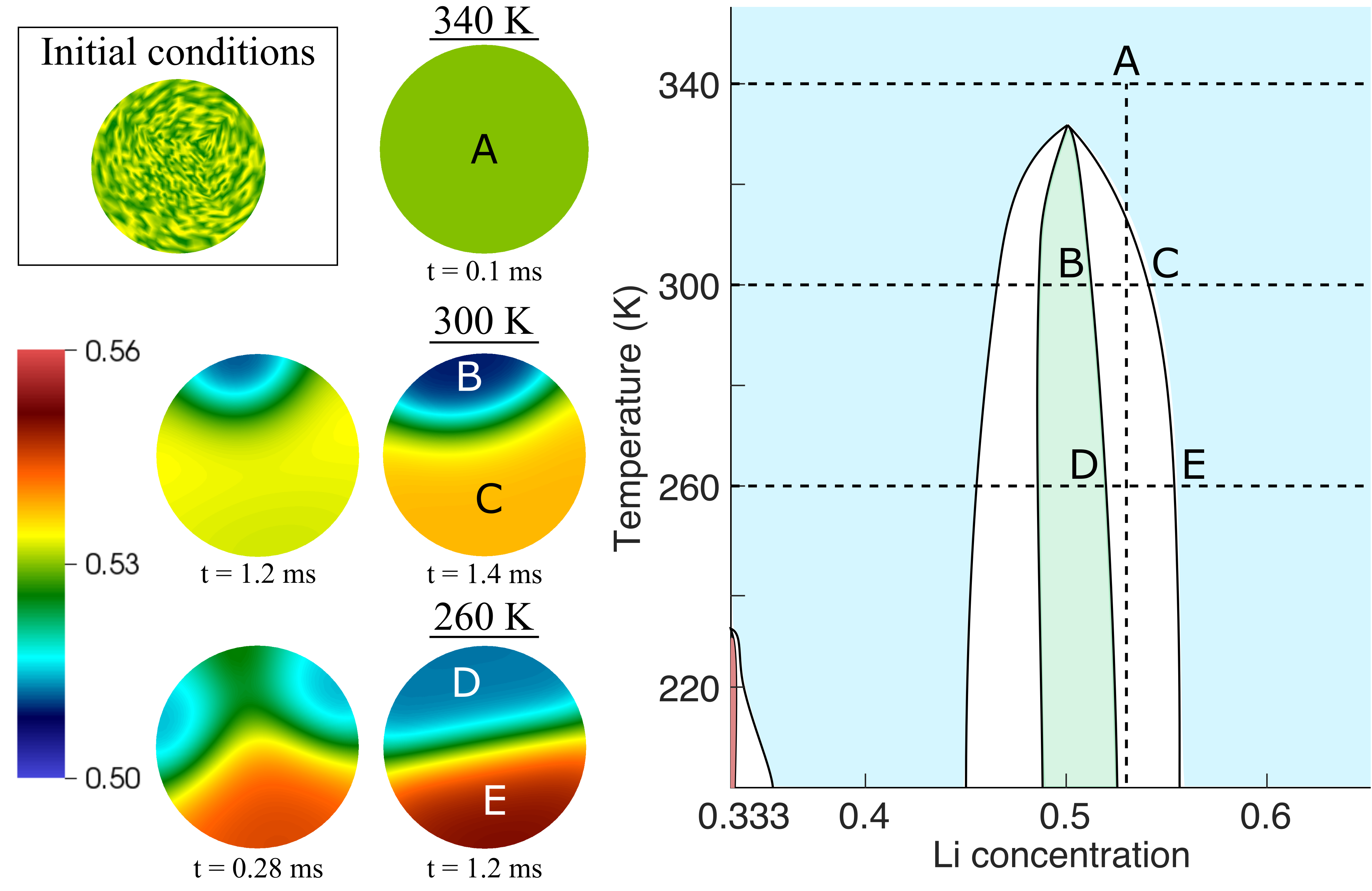}
\caption{2D phase field simulation results at 260, 300, and 340 K showing the Li composition in a 50 nm diameter particle, with initial Li composition randomly perturbed about $x = 0.53$ and no boundary flux. Compare with the results using 1 $\mu$m particles in Figure \ref{fig:spinodal}.}
\label{fig:spinodal_small}
\end{figure}

\begin{figure}[tb]
    \centering
    \includegraphics[width=0.7\textwidth]{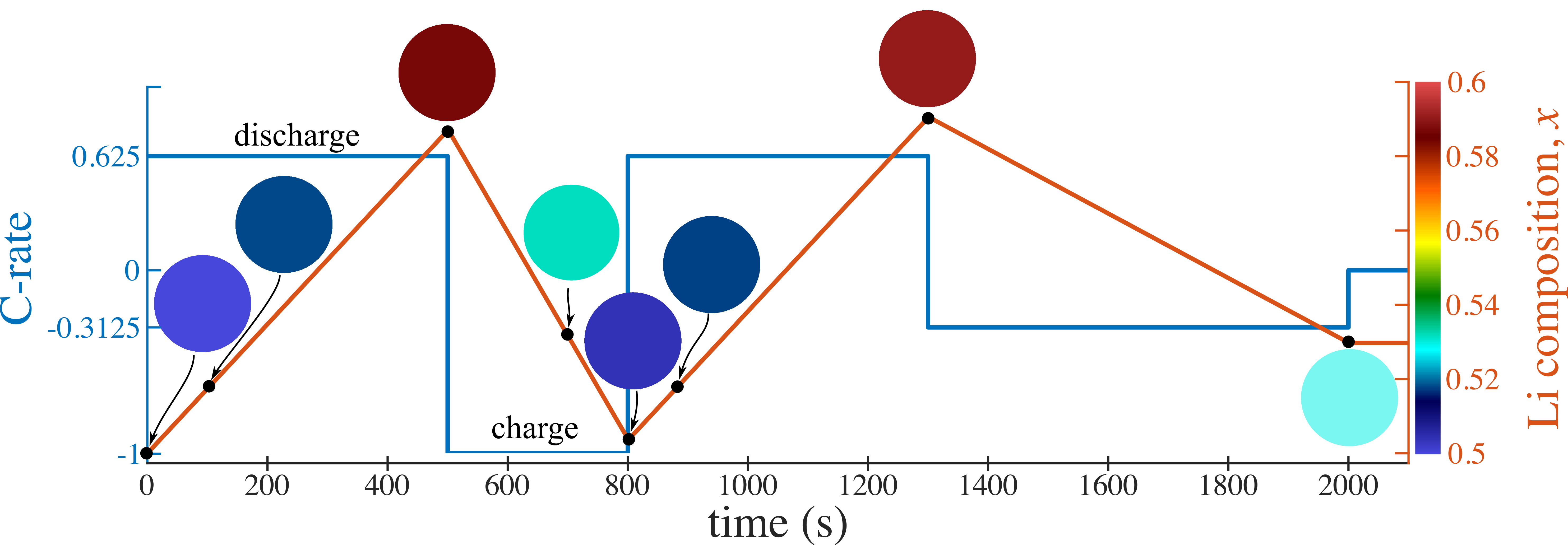}
\caption{Phase field results showing the Li composition field resulting from applying a cycling current density for a 50 nm diameter particle. Results are visually indistinguishable for 260, 300, and 340 K, and so only one set of results is included here. The applied current density is reduced by a factor of 20 compared with the 1 $\mu$m particles to maintain the same C-rates in both particle sizes (see Figure \ref{fig:cycle}). The blue line shows the applied C-rate, and the red line shows the corresponding average Li composition of the particle.}
\label{fig:cycle_small}
\end{figure}

While the phase field simulations in this work up to this point have used 1 $\mu$m diameter particles, experimental work by Choi et al. \cite{choi2006particle} has compared performance using micron sized LCO particles with that of 30-70 nm particles. Among other findings, they report that the cycle performance with these nanometer-sized particles is consistent, without deterioration, across their selected range of operating temperatures (258 to 333 K), and that the smaller sized particles lead to increase reliability. With this as motivation, spinodal decomposition and cycling phase field simulations were performed using a 50 nm diameter particle, again with the temperatures 260, 300, and 340 K. The current density applied in the cycling simulations was scaled down by a factor of 20 to maintain the same C-rates in the smaller particles as in the larger, micron sized particles. The same mobility functions and phase field gradient coefficient were used as with the larger particles.

Results from the phase field simulations initialized with Li composition randomly perturbed about $x = 0.53$, with no boundary flux, are shown in Figure \ref{fig:spinodal_small} for the 50 nm particle. As with the larger particles, spinodal decomposition occurs at 260 and 300 K, but not 340 K. However, equilibrium is reached between 100 and 10,000 times more quickly with the smaller particles, depending on the temperature. Additionally, while the equilibrium microstructures are comparable for both sizes, the transient microstructure is much simpler for the smaller particle. This is because the length scale of the transient microstructure in the simulations is defined by the gradient coefficient, which is related to the interfacial energy and is independent of the particle size. Since the length scale of the microstructure is maintained but the size of the particle is decreased, the result is a less complex microstructure. This reduced complexity could lead to diminished misfit stress effects and increased reliability.

The results from the cycling simulation for the 50 nm particles presented in Figure \ref{fig:cycle_small} show that, unlike with the larger particles, the Li composition field is essentially the same for all three temperatures. This is because the particles are small enough that the relative values of the mobility, particle size, and microstructure length scale lead the Li atoms to diffuse uniformly and extremely rapidly in comparison with the time scale of the simulation such that any temperature effects are not seen. This, again, matches the experimental findings of Choi et al. \cite{choi2006particle} which show consistent performance across the temperature range of 258 to 333 K.

\section{Discussion and conclusions}
\label{sec:conclusions}
In this work, we have presented a computational framework for modeling materials physics that bridges between atomic and continuum scales, as applied to the widely used battery material Li$_x$CoO$_2$. The framework begins with DFT calculations of the formation energy of potential configurations and compositions of Li$_x$CoO$_2$, bridging scales up through statistical mechanics Monte Carlo simulations and integrable deep neural network training, culminating in continuum phase field modeling. In particular, through the DFT and Monte Carlo simulations, the framework identified the order-disorder transition around composition $x=0.5$ at room temperature. This phase transition was subsequently captured within the neural network representation of the free energy and the formation and evolution of this transition was modeled in the phase field simulations. 

Our study focused on phase stability has been developed, thus far, for the O3 structure of LCO. However, since H1-3 form of LCO is the stable structure below $x = 1/3$, our results do not hold in that range. The corresponding DFT studies are underway, and the results from them will be subject to our scale-bridging framework to ultimately present a complete study that will have rigorous predictive value over the entire composition range. We note that, since LCO has two distinct free energy curves in the composition ranges $0 < x \le 1/3$ and $1/3 < x < 1$, a phase field treatment would not be thermodynamically rigorous over the entire composition range. Sharp interface methods such as levelsets present one appropriate framework in this setting.

With regard to charge splitting we note that our DFT+U calculations used the Hubbard correction with a van der Waals functional, and we highlighted three configurations at $x = 1/2$, which are reported in Figure \ref{fig:c=0.5}. Of these, the lowest energy configuration, which we adopted as the ground state, was found to be the zigzag ordering (configuration C in Figure \ref{fig:c=0.5}). Charge splitting was not observed for this ground state configuration, although we noted that  a more extensive investigation would require more advanced hybrid exchange functionals~\cite{b3lyp,pbe0,hse} or extended Hubbard models (DFT $+ U + V$)~\cite{Amaricci2010,campo2010extended}. If such detailed studies do predict charge splitting configurations, this information could be propagated up the scales by incorporating them in the cluster Hamiltonians and defining additional order parameters that would be identified from the Monte Carlo sampling. The active learning approach would play an important role in directing this sampling, which would take place in a much higher dimensional space to generate training data for the IDNN.

The choice of the $U$ parameter was guided by matching with experimental voltage measurements in the $x \in [0, 1/2]$ and $x \in [0,1]$ composition ranges (see Figure \ref{fig:volt_U}). This choice is finally reflected in the voltage reported across the entire $x \in [0,1]$ composition range in Figure \ref{fig:voltage}. We note that a different approach to calibration could be to choose $U$ for an optimal fit across the composition range in Figure \ref{fig:voltage}, although this would be a much more expensive and complex process that would have to fundamentally leverage the scale bridging between DFT and statistical mechanics. We further note that empirical approaches to determine the $U$ parameter can be completely circumvented by using the recently developed self-consistent Hubbard method~\cite{timrov2021} where the site specific values of $U$ are determined using density-functional perturbation theory and iterated self-consistently until convergence of the DFT $+ U$ ground-state. This approach seamlessly ties the $U$ values to the  changing composition through its dependence on the local chemical environment.

We re-emphasize that while the LCO material system has been the subject of a vast body of computational study at the atomic as well as at the mesoscopic continuum scale and thermodynamic investigations, all supported by experimental campaigns, the extent of scale bridging has not extended as widely as in this communication, to the best of our knowledge. Our choice of LCO has been motivated by this deep and broad literature, which provides a robust basis for comparison as well as the interest in this material system. More broadly,  this study also serves a demonstration of method development that will be applied to other interesting materials systems, with systematic incorporation of additional effects. On this last note, our first extension will be to the non-conserved symmetry-adapted order parameters, $\boldsymbol{\eta}$ discussed in Section \ref{sec:order parameters}. Other extensions are also possible to coupling the phase field description with a thermodynamically rigorous treatment of heat generation and  transport that will more completely reflect the phase diagram. Scale-bridged treatments of strain effects, misfit elasticity, charge splitting as well as the dependence of coefficients on all these fields are also accessible. An extension to the equations of electrochemistry at the active particle-electrolyte scale will allow the study of phenomena such as the mosaic instability observed in nanoparticulate batteries, wherein phase segregation occurs between distinct particles rather than within a single particle \cite{orvananos2014particle} as studied here. Even without such a level of rigorously obtained detailed physics, we note that the phase field treatment exposes trends that both align with battery material particle and cell scale experiments, and present a platform for making technologically relevant projections as discussed in Section \ref{sec:phasefieldresults}.

\section*{Author contributions}
\label{sec:contributions}
The study's scale bridging methodology was developed by KG and GHT; electronic structure studies were planned by all the authors;  electronic structure calculations were performed by GHT and SD; statistical mechanics and phase field studies were carried out by GHT; the paper was written by all the authors.

\section*{Acknowledgements}
\label{sec:acknowledgements}
We thank Brian Puchala for his insight regarding the \texttt{CASM} software and related methods. We also gratefully acknowledge the support of Toyota Research Institute, Award \#849910, ``Computational framework for data-driven, predictive, multi-scale and multi-physics modeling of battery materials''.
We acknowledge the support of the U.S. Army Research Office through the DURIP grant W911NF1810242, which provided the computational resources for this work.
This work has also been supported in part by National Science Foundation DMREF grant \#1729166, ``Integrated Framework for Design of Alloy-Oxide Structures''.
Additional support was provided by Defense Advanced Research Projects Agency (DARPA) under Agreement No. HR0011199002, ``Artificial Intelligence guided multi-scale multi-physics framework for discovering complex emergent materials phenomena''.
Simulations in this work were also performed in part using resources were provided by the Extreme Science and Engineering Discovery Environment (XSEDE) Comet at the San Diego Supercomputer Center and Stampede2 at the Texas Advanced Computing Center through allocations TG-DMR180072 and TG-MCH200011.
The phase field simulations were performed using resources provided by the NSF via grant 1531752 MRI: Acquisition of Conflux, A Novel Platform for Data-Driven Computational Physics (Tech. Monitor: Ed Walker), with past and current support by the University of Michigan.
XSEDE is supported by National Science Foundation grant number ACI-1548562.

\bibliographystyle{unsrt}
\bibliography{references}

\begin{thebibliography}{10}

\bibitem{Zheng2019}
Shiyao Zheng, Chaoyu Hong, Xiaoyun Guan, Yuxuan Xiang, Xiangsi Liu, Gui-Liang
  Xu, Rui Liu, Guiming Zhong, Feng Zheng, Yixiao Li, Xiaoyi Zhang, Yang Ren,
  Zonghai Chen, Khalil Amine, and Yong Yang.
\newblock Correlation between long range and local structural changes in
  ni-rich layered materials during charge and discharge process.
\newblock {\em Journal of Power Sources}, 412:336--343, 2019.

\bibitem{Mu2018}
Linqin Mu, Qingxi Yuan, Chixia Tian, Chenxi Wei, Kai Zhang, Jin Liu, Piero
  Pianetta, Marca~M. Doeff, Yijin Liu, and Feng Lin.
\newblock Propagation topography of redox phase transformations in
  heterogeneous layered oxide cathode materials.
\newblock {\em Nature Communications}, 9(1):2810, Jul 2018.

\bibitem{Yang2019}
Yang Yang, Rong Xu, Kai Zhang, Sang-Jun Lee, Linqin Mu, Pengfei Liu, Crystal~K.
  Waters, Stephanie Spence, Zhengrui Xu, Chenxi Wei, David~J. Kautz, Qingxi
  Yuan, Yuhui Dong, Young-Sang Yu, Xianghui Xiao, Han-Koo Lee, Piero Pianetta,
  Peter Cloetens, Jun-Sik Lee, Kejie Zhao, Feng Lin, and Yijin Liu.
\newblock Quantification of heterogeneous degradation in li-ion batteries.
\newblock {\em Advanced Energy Materials}, 9(25):1900674, 2019.

\bibitem{Mao2019}
Yuwei Mao, Xuelong Wang, Sihao Xia, Kai Zhang, Chenxi Wei, Seongmin Bak,
  Zulipiya Shadike, Xuejun Liu, Yang Yang, Rong Xu, Piero Pianetta, Stefano
  Ermon, Eli Stavitski, Kejie Zhao, Zhengrui Xu, Feng Lin, Xiao-Qing Yang,
  Enyuan Hu, and Yijin Liu.
\newblock High-voltage charging-induced strain, heterogeneity, and micro-cracks
  in secondary particles of a nickel-rich layered cathode material.
\newblock {\em Advanced Functional Materials}, 29(18):1900247, 2019.

\bibitem{Liu2015}
Wen Liu, Pilgun Oh, Xien Liu, Min-Joon Lee, Woongrae Cho, Sujong Chae, Youngsik
  Kim, and Jaephil Cho.
\newblock Nickel-rich layered lithium transition-metal oxide for high-energy
  lithium-ion batteries.
\newblock {\em Angewandte Chemie International Edition}, 54(15):4440--4457,
  2015.

\bibitem{Hou2017}
Peiyu Hou, Jiangmei Yin, Meng Ding, Jinzhao Huang, and Xijin Xu.
\newblock Surface/interfacial structure and chemistry of high-energy
  nickel-rich layered oxide cathodes: Advances and perspectives.
\newblock {\em Small}, 13(45):1701802, 2017.

\bibitem{Kim2017}
Un-Hyuck Kim, Seung-Taek Myung, Chong~S. Yoon, and Yang-Kook Sun.
\newblock Extending the battery life using an al-doped li[ni0.76co0.09mn0.15]o2
  cathode with concentration gradients for lithium ion batteries.
\newblock {\em ACS Energy Letters}, 2(8):1848--1854, 2017.

\bibitem{Kong2019}
Defei Kong, Jiangtao Hu, Zhefeng Chen, Kepeng Song, Cheng Li, Mouyi Weng,
  Maofan Li, Rui Wang, Tongchao Liu, Jiajie Liu, Mingjian Zhang, Yinguo Xiao,
  and Feng Pan.
\newblock Ti-gradient doping to stabilize layered surface structure for high
  performance high-ni oxide cathode of li-ion battery.
\newblock {\em Advanced Energy Materials}, 9(41):1901756, 2019.

\bibitem{Weigel2019}
Tina Weigel, Florian Schipper, Evan~M. Erickson, Francis~Amalraj Susai, Boris
  Markovsky, and Doron Aurbach.
\newblock Structural and electrochemical aspects of lini0.8co0.1mn0.1o2 cathode
  materials doped by various cations.
\newblock {\em ACS Energy Letters}, 4(2):508--516, 2019.

\bibitem{Teichert2019}
G.H. Teichert, A.R. Natarajan, A.~Van der Ven, and K.~Garikipati.
\newblock Machine learning materials physics: Integrable deep neural networks
  enable scale bridging by learning free energy functions.
\newblock {\em Computer Methods in Applied Mechanics and Engineering}, 353:201
  -- 216, 2019.

\bibitem{Teichert2020}
G.H. Teichert, A.R. Natarajan, A.~{Van der Ven}, and K.~Garikipati.
\newblock Scale bridging materials physics: Active learning workflows and
  integrable deep neural networks for free energy function representations in
  alloys.
\newblock {\em Computer Methods in Applied Mechanics and Engineering},
  371:113281, 2020.

\bibitem{Reimers1992}
Jan~N. Reimers and J.~R. Dahn.
\newblock Electrochemical and in situ x-ray diffraction studies of lithium
  intercalation in {Li}$_x${CoO}$_2$.
\newblock {\em Journal of The Electrochemical Society}, 139(8):2091--2097, Aug
  1992.

\bibitem{Menetrier1999}
M.~Ménétrier, I.~Saadoune, S.~Levasseur, and C.~Delmas.
\newblock The insulator-metal transition upon lithium deintercalation from
  {LiCoO$_2$}: Electronic properties and {7Li NMR} study.
\newblock {\em Journal of Materials Chemistry}, 9(5):1135--1140, 1999.
\newblock cited By 337.

\bibitem{ShaoHorn2003}
Y.~Shao-Horn, S.~Levasseur, F.~Weill, and C.~Delmas.
\newblock Probing lithium and vacancy ordering in {O3} layered
  {Li}$_x${CoO}$_2$ (x$\approx$0.5).
\newblock {\em Journal of The Electrochemical Society}, 150(3):A366, 2003.

\bibitem{Takahashi2007}
Y~Takahashi, N~Kijima, K~Tokiwa, T~Watanabe, and J~Akimoto.
\newblock Single-crystal synthesis, structure refinement and electrical
  properties of {Li}$_{0.5}${CoO}$_2$.
\newblock {\em Journal of Physics: Condensed Matter}, 19(43):436202, Sep 2007.

\bibitem{Motohashi2009}
T.~Motohashi, T.~Ono, Y.~Sugimoto, Y.~Masubuchi, S.~Kikkawa, R.~Kanno,
  M.~Karppinen, and H.~Yamauchi.
\newblock Electronic phase diagram of the layered cobalt oxide system
  {Li$_x$CoO$_2$} $(0.0\ensuremath{\le}x\ensuremath{\le}1.0)$.
\newblock {\em Phys. Rev. B}, 80:165114, Oct 2009.

\bibitem{Chen2004}
Zhaohui Chen and J.R. Dahn.
\newblock Methods to obtain excellent capacity retention in {LiCoO$_2$} cycled
  to 4.5 v.
\newblock {\em Electrochimica Acta}, 49(7):1079 -- 1090, 2004.

\bibitem{Chang2013}
Keke Chang, Bengt Hallstedt, Denis Music, Julian Fischer, Carlos Ziebert, Sven
  Ulrich, and Hans~J. Seifert.
\newblock Thermodynamic description of the layered {O3} and {O2} structural
  {LiCoO$_2$–CoO$_2$} pseudo-binary systems.
\newblock {\em Calphad}, 41:6 -- 15, 2013.

\bibitem{VanderVen1998}
A.~Van~der Ven, M.~K.. Aydinol, G.~Ceder, G.~Kresse, and J.~Hafner.
\newblock First-principles investigation of phase stability in {Li$_x$CoO$_2$}.
\newblock {\em Phys. Rev. B}, 58:2975--2987, 1998.

\bibitem{Wolverton1998}
C.~Wolverton and Alex Zunger.
\newblock First-principles prediction of vacancy order-disorder and
  intercalation battery voltages in {Li$_x$CoO$_2$}.
\newblock {\em Phys. Rev. Lett.}, 81:606--609, Jul 1998.

\bibitem{Abe2011}
Taichi Abe and Toshiyuki Koyama.
\newblock Thermodynamic modeling of the {LiCoO$_2$}–{CoO$_2$} pseudo-binary
  system.
\newblock {\em Calphad}, 35(2):209 -- 218, 2011.

\bibitem{Nadkarni2019}
Neel Nadkarni, Tingtao Zhou, Dimitrios Fraggedakis, Tao Gao, and Martin~Z.
  Bazant.
\newblock Modeling the metal–insulator phase transition in {LixCoO$_2$} for
  energy and information storage.
\newblock {\em Advanced Functional Materials}, 29(40):1902821, 2019.

\bibitem{Cococcioni2005}
Matteo Cococcioni and Stefano de~Gironcoli.
\newblock Linear response approach to the calculation of the effective
  interaction parameters in the $\mathrm{LDA}+\mathrm{U}$ method.
\newblock {\em Phys. Rev. B}, 71:035105, Jan 2005.

\bibitem{Aykol2015}
Muratahan Aykol, Soo Kim, and C.~Wolverton.
\newblock van der waals interactions in layered lithium cobalt oxides.
\newblock {\em The Journal of Physical Chemistry C}, 119(33):19053--19058,
  2015.

\bibitem{PhysRevB.86.134117}
Chirranjeevi~Balaji Gopal and Axel van~de Walle.
\newblock Ab initio thermodynamics of intrinsic oxygen vacancies in ceria.
\newblock {\em Phys. Rev. B}, 86:134117, Oct 2012.

\bibitem{y2002first}
ME~Arroyo y~de Dompablo, A~Van~der Ven, and G~Ceder.
\newblock First-principles calculations of lithium ordering and phase stability
  on li x nio 2.
\newblock {\em Physical Review B}, 66(6):064112, 2002.

\bibitem{Anisimov1991}
Vladimir~I. Anisimov, Jan Zaanen, and Ole~K. Andersen.
\newblock {Band theory and Mott insulators: Hubbard U instead of Stoner I}.
\newblock {\em Phys. Rev. B}, 44:943--954, Jul 1991.

\bibitem{Zhou2004}
F.~Zhou, M.~Cococcioni, C.~A. Marianetti, D.~Morgan, and G.~Ceder.
\newblock First-principles prediction of redox potentials in transition-metal
  compounds with {$\mathrm{LDA}+U$}.
\newblock {\em Phys. Rev. B}, 70:235121, Dec 2004.

\bibitem{Aykol2014}
Muratahan Aykol and C.~Wolverton.
\newblock Local environment dependent $\text{GGA}+u$ method for accurate
  thermochemistry of transition metal compounds.
\newblock {\em Phys. Rev. B}, 90:115105, Sep 2014.

\bibitem{Thonhauser2007}
T.~Thonhauser, Valentino~R. Cooper, Shen Li, Aaron Puzder, Per Hyldgaard, and
  David~C. Langreth.
\newblock {Van der Waals density functional: Self-consistent potential and the
  nature of the van der Waals bond}.
\newblock {\em Phys. Rev. B}, 76:125112, Sep 2007.

\bibitem{Klime2009}
Ji{\v{r}}{\'{\i}} Klime{\v{s}}, David~R Bowler, and Angelos Michaelides.
\newblock {Chemical accuracy for the van der Waals density functional}.
\newblock {\em Journal of Physics: Condensed Matter}, 22(2):022201, Dec 2009.

\bibitem{Langreth2009}
D~C Langreth, B~I Lundqvist, S~D Chakarova-Käck, V~R Cooper, M~Dion,
  P~Hyldgaard, A~Kelkkanen, J~Kleis, Lingzhu Kong, Shen Li, P~G Moses,
  E~Murray, A~Puzder, H~Rydberg, E~Schröder, and T~Thonhauser.
\newblock A density functional for sparse matter.
\newblock {\em Journal of Physics: Condensed Matter}, 21(8):084203, Jan 2009.

\bibitem{Sabatini2012}
Riccardo Sabatini, Emine Kü{\c{c}}ükbenli, Brian Kolb, T~Thonhauser, and
  Stefano de~Gironcoli.
\newblock Structural evolution of amino acid crystals under stress from a
  non-empirical density functional.
\newblock {\em Journal of Physics: Condensed Matter}, 24(42):424209, Oct 2012.

\bibitem{Thonhauser2015}
T.~Thonhauser, S.~Zuluaga, C.~A. Arter, K.~Berland, E.~Schr\"oder, and
  P.~Hyldgaard.
\newblock Spin signature of nonlocal correlation binding in metal-organic
  frameworks.
\newblock {\em Phys. Rev. Lett.}, 115:136402, Sep 2015.

\bibitem{Berland2015}
Kristian Berland, Valentino~R Cooper, Kyuho Lee, Elsebeth Schröder,
  T~Thonhauser, Per Hyldgaard, and Bengt~I Lundqvist.
\newblock van der waals forces in density functional theory: a review of the
  {vdW}-{DF} method.
\newblock {\em Reports on Progress in Physics}, 78(6):066501, May 2015.

\bibitem{QE-2009}
Paolo Giannozzi, Stefano Baroni, Nicola Bonini, Matteo Calandra, Roberto Car,
  Carlo Cavazzoni, Davide Ceresoli, Guido~L Chiarotti, Matteo Cococcioni,
  Ismaila Dabo, Andrea {Dal Corso}, Stefano de~Gironcoli, Stefano Fabris, Guido
  Fratesi, Ralph Gebauer, Uwe Gerstmann, Christos Gougoussis, Anton Kokalj,
  Michele Lazzeri, Layla Martin-Samos, Nicola Marzari, Francesco Mauri,
  Riccardo Mazzarello, Stefano Paolini, Alfredo Pasquarello, Lorenzo Paulatto,
  Carlo Sbraccia, Sandro Scandolo, Gabriele Sclauzero, Ari~P Seitsonen,
  Alexander Smogunov, Paolo Umari, and Renata~M Wentzcovitch.
\newblock Quantum espresso: a modular and open-source software project for
  quantum simulations of materials.
\newblock {\em Journal of Physics: Condensed Matter}, 21(39):395502 (19pp),
  2009.

\bibitem{QE-2017}
P~Giannozzi, O~Andreussi, T~Brumme, O~Bunau, M~Buongiorno Nardelli, M~Calandra,
  R~Car, C~Cavazzoni, D~Ceresoli, M~Cococcioni, N~Colonna, I~Carnimeo, A~Dal
  Corso, S~de~Gironcoli, P~Delugas, R~A~DiStasio Jr, A~Ferretti, A~Floris,
  G~Fratesi, G~Fugallo, R~Gebauer, U~Gerstmann, F~Giustino, T~Gorni, J~Jia,
  M~Kawamura, H-Y Ko, A~Kokalj, E~Küçükbenli, M~Lazzeri, M~Marsili,
  N~Marzari, F~Mauri, N~L Nguyen, H-V Nguyen, A~Otero de-la Roza, L~Paulatto,
  S~Poncé, D~Rocca, R~Sabatini, B~Santra, M~Schlipf, A~P Seitsonen,
  A~Smogunov, I~Timrov, T~Thonhauser, P~Umari, N~Vast, X~Wu, and S~Baroni.
\newblock Advanced capabilities for materials modelling with quantum espresso.
\newblock {\em Journal of Physics: Condensed Matter}, 29(46):465901, 2017.

\bibitem{dalcorso2014}
Andrea {Dal Corso}.
\newblock Pseudopotentials periodic table: From {H} to {Pu}.
\newblock {\em Computational Materials Science}, 95:337 -- 350, 2014.

\bibitem{pslibrary}
Pslibrary 1.0.0.
\newblock {https://github.com/dalcorso/pslibrary}.

\bibitem{Momma2011}
K.~Momma and F.~Izumi.
\newblock {VESTA} 3 for three-dimensional visualization of crystal, volumetric
  and morphology data.
\newblock {\em J. Appl. Crystallogr.}, 44:1272--1276, 2011.

\bibitem{Meredig2010}
B.~Meredig, A.~Thompson, H.~A. Hansen, C.~Wolverton, and A.~van~de Walle.
\newblock Method for locating low-energy solutions within $\text{DFT}+u$.
\newblock {\em Phys. Rev. B}, 82:195128, Nov 2010.

\bibitem{Aydinol1997}
M.~K. Aydinol, A.~F. Kohan, G.~Ceder, K.~Cho, and J.~Joannopoulos.
\newblock Ab initio study of lithium intercalation in metal oxides and metal
  dichalcogenides.
\newblock {\em Phys. Rev. B}, 56:1354--1365, Jul 1997.

\bibitem{Natarajan2017}
Anirudh~Raju Natarajan, John~C. Thomas, Brian Puchala, and Anton Van~der Ven.
\newblock Symmetry-adapted order parameters and free energies for solids
  undergoing order-disorder phase transitions.
\newblock {\em Phys. Rev. B}, 96:134204, Oct 2017.

\bibitem{Sanchez1984}
J.~M. Sanchez, F.~Ducastelle, and D.~Gratias.
\newblock Generalized cluster description of multicomponent systems.
\newblock {\em Physica A}, 128:334--350, 1984.

\bibitem{deFontaine1994}
D.~De Fontaine.
\newblock Cluster approach to order-disorder transformations in alloys.
\newblock volume~47 of {\em Solid State Physics}, pages 33 -- 176. Academic
  Press, 1994.

\bibitem{casm}
\texttt{https://github.com/prisms-center/CASMcode}.
\newblock {{CASM}}: {{A Clusters Approach}} to {{Statistical Mechanics}},
  v0.3.dev, 2018.

\bibitem{VanderVen2010}
A.~Van~der Ven, J.~C. Thomas, Q.~Xu, and J.~Bhattacharya.
\newblock Linking the electronic structure of solids to their thermodynamic and
  kinetic properties.
\newblock {\em Mathematics and Computers in Simulation}, 80(7):1393--1410,
  2010.

\bibitem{thomas2013}
John~C. Thomas and Anton {Van der Ven}.
\newblock Finite-temperature properties of strongly anharmonic and mechanically
  unstable crystal phases from first principles.
\newblock {\em Physical Review B}, 88(21):214111--214111, December 2013.

\bibitem{puchala2013}
B.~Puchala and A.~Van~der Ven.
\newblock Thermodynamics of the {Zr-O} system from first-principles
  calculations.
\newblock {\em Phys. Rev. B}, 88:094108, Sep 2013.

\bibitem{mishin2004}
Y.~Mishin.
\newblock Atomistic modeling of the $\gamma$ and $\gamma^\prime$-phases of the
  {Ni–Al} system.
\newblock {\em Acta Materialia}, 52(6):1451 -- 1467, 2004.

\bibitem{Sadigh2012a}
Babak Sadigh, Paul Erhart, Alexander Stukowski, Alfredo Caro, Enrique Martinez,
  and Luis Zepeda-Ruiz.
\newblock Scalable parallel monte carlo algorithm for atomistic simulations of
  precipitation in alloys.
\newblock {\em Phys. Rev. B}, 85:184203, May 2012.

\bibitem{Sadigh2012b}
Babak Sadigh and Paul Erhart.
\newblock Calculation of excess free energies of precipitates via direct
  thermodynamic integration across phase boundaries.
\newblock {\em Phys. Rev. B}, 86:134204, Oct 2012.

\bibitem{thomas2017}
John~C. Thomas and Anton {Van der Ven}.
\newblock The exploration of nonlinear elasticity and its efficient
  parameterization for crystalline materials.
\newblock {\em Journal of the Mechanics and Physics of Solids}, 107:76--95,
  October 2017.

\bibitem{CahnHilliard1958}
J.~W. Cahn and J.~E. Hilliard.
\newblock Free energy of a nonuniform system. {I} {I}nterfacial energy.
\newblock {\em The Journal of Chemical Physics}, 28:258--267, 1958.

\bibitem{Allen1979}
S.~M. Allen and J.~W. Cahn.
\newblock A microscopic theory for antiphase boundary motion and its
  application to antiphase boundary coarsening.
\newblock {\em Acta Metallurgica}, 27:1085–1091, 1979.

\bibitem{VanderVen2000}
A.~Van der Ven.
\newblock Lithium diffusion in layered {Li$_x$CoO$_2$}.
\newblock {\em Electrochemical and Solid-State Letters}, 3(7):301, 2000.

\bibitem{Jiang2016}
Tonghu Jiang, Shiva Rudraraju, Anindya Roy, Anton Van~der Ven, Krishna
  Garikipati, and Michael~L. Falk.
\newblock Multiphysics simulations of lithiation-induced stress in
  {Li$_{1+x}$Ti$_2$O$_4$} electrode particles.
\newblock {\em The Journal of Physical Chemistry C}, 120(49):27871--27881,
  2016.

\bibitem{dealii2017}
Daniel Arndt, Wolfgang Bangerth, Denis Davydov, Timo Heister, Luca Heltai,
  Martin Kronbichler, Matthias Maier, Jean-Paul Pelteret, Bruno Turcksin, and
  David Wells.
\newblock The deal.{II} library, version 8.5.
\newblock {\em Journal of Numerical Mathematics}, 25(3), April 2017.

\bibitem{b3lyp}
Axel~D. Becke.
\newblock {Density‐functional thermochemistry. III. The role of exact
  exchange}.
\newblock {\em The Journal of Chemical Physics}, 98(7):5648--5652, 1993.

\bibitem{pbe0}
John~P. Perdew, Matthias Ernzerhof, and Kieron Burke.
\newblock Rationale for mixing exact exchange with density functional
  approximations.
\newblock {\em The Journal of Chemical Physics}, 105(22):9982--9985, 1996.

\bibitem{hse}
Jochen Heyd, Gustavo~E. Scuseria, and Matthias Ernzerhof.
\newblock Hybrid functionals based on a screened coulomb potential.
\newblock {\em The Journal of Chemical Physics}, 118(18):8207--8215, 2003.

\bibitem{Amaricci2010}
A.~Amaricci, A.~Camjayi, K.~Haule, G.~Kotliar,
  D.~Tanaskovi\ifmmode~\acute{c}\else \'{c}\fi{}, and
  V.~Dobrosavljevi\ifmmode~\acute{c}\else \'{c}\fi{}.
\newblock Extended hubbard model: Charge ordering and wigner-mott transition.
\newblock {\em Phys. Rev. B}, 82:155102, Oct 2010.

\bibitem{campo2010extended}
Vivaldo~Leiria Campo~Jr and Matteo Cococcioni.
\newblock Extended {DFT+ U+ V} method with on-site and inter-site electronic
  interactions.
\newblock {\em Journal of Physics: Condensed Matter}, 22(5):055602, 2010.

\bibitem{Amatucci1996}
G.~G. Amatucci, J.~M. Tarascon, and L.~C. Klein.
\newblock {CoO}$_2$, the end member of the {Li}$_x${CoO}$_2$ solid solution.
\newblock {\em Journal of The Electrochemical Society}, 143(3):1114--1123, Mar
  1996.

\bibitem{wang1999tem}
Haifeng Wang, Young-Il Jang, Biying Huang, Donald~R Sadoway, and Yet-Ming
  Chiang.
\newblock Tem study of electrochemical cycling-induced damage and disorder in
  licoo2 cathodes for rechargeable lithium batteries.
\newblock {\em Journal of the Electrochemical Society}, 146(2):473, 1999.

\bibitem{choi2006particle}
Sun~Hee Choi, Ji-Won Son, Young~Soo Yoon, and Joosun Kim.
\newblock Particle size effects on temperature-dependent performance of licoo2
  in lithium batteries.
\newblock {\em Journal of power sources}, 158(2):1419--1424, 2006.

\bibitem{pender2020electrode}
Joshua~P Pender, Gaurav Jha, Duck~Hyun Youn, Joshua~M Ziegler, Ilektra Andoni,
  Eric~J Choi, Adam Heller, Bruce~S Dunn, Paul~S Weiss, Reginald~M Penner,
  et~al.
\newblock Electrode degradation in lithium-ion batteries.
\newblock {\em ACS nano}, 14(2):1243--1295, 2020.

\bibitem{leng2015effect}
Feng Leng, Cher~Ming Tan, and Michael Pecht.
\newblock Effect of temperature on the aging rate of li ion battery operating
  above room temperature.
\newblock {\em Scientific reports}, 5(1):1--12, 2015.

\bibitem{timrov2021}
Iurii Timrov, Nicola Marzari, and Matteo Cococcioni.
\newblock Self-consistent hubbard parameters from density-functional
  perturbation theory in the ultrasoft and projector-augmented wave
  formulations.
\newblock {\em Phys. Rev. B}, 103:045141, Jan 2021.

\bibitem{orvananos2014particle}
Bernardo Orvananos, Todd~R Ferguson, Hui-Chia Yu, Martin~Z Bazant, and Katsuyo
  Thornton.
\newblock Particle-level modeling of the charge-discharge behavior of
  nanoparticulate phase-separating li-ion battery electrodes.
\newblock {\em Journal of The Electrochemical Society}, 161(4):A535, 2014.

\end{thebibliography}

\end{document}